\documentclass[11pt,a4paper]{article}
\usepackage{jstyle}

\newcommand{\bi}{\begin{itemize}}
\newcommand{\ei}{\end{itemize}}
\newcommand{\bea}{\begin{align}}
\newcommand{\eea}{\end{align}}
\newcommand{\be}{\begin{equation}}
\newcommand{\ee}{\end{equation}}

\newcommand{\ga}{\alpha}

\newcommand{\pl}{{\partial}}

\makeatletter
\renewcommand*\env@matrix[1][\arraystretch]{%
  \edef\arraystretch{#1}%
  \hskip -\arraycolsep
  \let\@ifnextchar\new@ifnextchar
  \array{*\c@MaxMatrixCols c}}
\makeatother

\author{\quad Massimo TARONNA$^{a}$ \footnote{Postdoctoral Researcher of the Fund for Scientific Research-FNRS Belgium.}}

\affiliation{${}^{a}$Physique Th\'eorique et Math\'ematique\\
\hspace*{0.01cm} Universit\'e Libre de Bruxelles
and International Solvay Institutes\\
\hspace*{0.01cm} ULB-Campus Plaine CP231, 1050 Brussels, Belgium}

\emailAdd{massimo.taronna@ulb.ac.be}

\title{\centering
\LARGE{
On the Non-Local Obstruction to\\ Interacting Higher Spins in Flat Space
}}

\abstract{Owing to a renewed interest in flat space higher spin gauge theories, in this note we provide further details and clarifications on the results presented in arXiv:1107.5843 and arXiv: 1209.5755, which investigated their locality properties. Focusing, for simplicity, on quartic couplings with one of the external legs having non-zero integer spin (which can be considered as a prototype for Weinberg-type arguments), we review the appearance of $1/\Box$ non-localities. In particular, we emphasise that it appears to be not possible to eliminate all of the aforementioned non-localities in the general quartic Noether procedure solution with a judicious choice of coupling constants and spectrum.
We also discuss the light-cone gauge fixing in $d=4$, and argue that the non-local obstruction discussed in the covariant language cannot be avoided using light-cone gauge formalism.}

\begin{document}

\maketitle

\section{Introduction}

The interaction problem in higher-spin (HS) gauge theories on flat space has garnered a re-newed interest in the last year \cite{Bengtsson:2016jfk,Bengtsson:2016hss,Ponomarev:2016jqk,Conde:2016izb,Sleight:2016xqq,Ponomarev:2016lrm,Ponomarev:2016cwi,Haehnel:2016mlb,Adamo:2016ple,Roiban:2017iqg}. These efforts have drawn inspiration from Metsaev's result \cite{Metsaev:1991mt,Metsaev:1991nb} for flat space cubic couplings, obtained using light-cone methods \cite{Bengtsson:1983pd,Bengtsson:1986kh} by solving for Poincar\'e invariance at quartic order. On the other hand, it is well known that interactions of HS gauge fields on flat space are threatened by a number of no-go results, most notably Weinberg theorem \cite{Weinberg:1964ew} (see \cite{Bekaert:2010hw,Rahman:2015pzl} for some reviews on the subject). The issue of the non-locality of HS interactions also plays a key role, and is the focus of the present note. Therefore, in the light of aforementioned new works on the subject, we take the opportunity to review, and reiterate, some of the statements and results of the previous works \cite{Taronna:2011kt,Taronna:2012gb} on non-local obstructions to interacting HS theories in flat space.\footnote{See also \cite{Sagnotti:2010at,Leclercq:2010ax,Bekaert:2010hp,Dempster:2012vw,Joung:2013nma} for some related results employing different techniques and assumptions.}

Locality, or more generally a functional space of permitted (admissible) non-localities, is a crucial property, and must be prescribed for predictability of the Noether procedure in a field theory context. If no locality condition is imposed, then the Noether procedure can be formally solved up to any order and for any choice of couplings \cite{Barnich:1993vg}. Moreover, this leads to the ability to perform arbitrary non-local field redefinitions which map, order by order in a weak field expansion, any interaction to zero. This would leave only free theories among the possible HS gauge field theories.\footnote{The issue of functional class and locality has also acquired some attention in AdS (see e.g. \cite{Prokushkin:1999xq,Bekaert:2014cea,Vasiliev:2015wma,Kessel:2015kna,Boulanger:2015ova,Bekaert:2015tva,Skvortsov:2015lja,Taronna:2016ats,Bekaert:2016ezc,Taronna:2016xrm,Sleight:2016hyl,Sleight:2017krf}). The full list of local cubic couplings in the type A theory were recently obtained in \cite{Sleight:2016dba}, via the holographic reconstruction program (see e.g. \cite{Sleight:2017krf} for a short review on this topic).}

The main issue we expound and review in this article is the non-local obstruction to HS interactions in flat space at quartic order, which was originally discussed in \cite{Taronna:2011kt}. In particular, under certain well-motivated assumptions on the convergence of summations over derivatives and spins, no solution to the Noether procedure without non-admissible $1/\Box$ singularities can be constructed for quartic couplings with higher-spin external legs. We discuss how \emph{some} obstructions may be removed by setting certain couplings to zero\footnote{Such as the scalar self-interactions.} (in accordance with Weinberg theorem\footnote{See also \cite{Porrati:2008rm} for related results.} \cite{Weinberg:1964ew}) or by enlarging the spectrum, but in spite of this $1/\Box$ obstructions still remain.

While the above results were obtained maintaining manifest Lorentz covariance, we also discuss the problem in the light-cone formalism. We show that Metsaev's cubic couplings can be recovered choosing the coupling constants such that the maximum number of non-localities are cancelled at quartic order in the Noether procedure. 
We then discuss the implication of Weinberg's theorem, concluding that a necessary condition to remove the remaining non-local obstructions in four-dimensions on the light-cone is the existence of additional Poincar\'e invariant structures on top of the standard structures which exists in the covariant classification. We discuss whether or not the existence of the aforementioned extra structures is feasible.

This note is organised as follows: Section \ref{noether} gives a brief introduction to the Noether procedure. Section \ref{Homosol} discusses the most general homogeneous solution to Noether procedure in the case where there is only one external spinning leg. Section \ref{loc} discusses the issue of locality and the obstructions arising in covariant formalism together with a list of caveats. Section \ref{lightcone} is instead devoted to the light-cone gauge and to the relation between the result obtained in covariant language and the result obtained by Metsaev using light-cone consistency in \cite{Metsaev:1991mt,Metsaev:1991nb}. In the appendices we discuss conventions \ref{not}, review the cubic coupling classification \ref{cubic}, some supplementary material like the Yang-Mills example \ref{YMnoether}, the computation of general exchange amplitudes both in 4d \ref{currexch} and generic dimension \ref{currexchd} and also more general couplings of the type $s_1$-$s_2$-$s_3$-$s_4$ and the type of non-local obstructions arising in the general case \ref{s1s1}. In particular we show that, at fixed external spins $s_i$, the non-local obstruction comes only from a sub-sector of exchange amplitudes with spin bounded from above, while the remainder of the sum over spins does not generate obstructions of this type and should be analysed at the next order or in relation to higher-spin symmetry (see e.g. \cite{Sleight:2016xqq}).  

\section{Noether procedure}\label{noether}

Noether procedure is a systematic scheme with the purpose to find interactions compatible with the deformation of the linear gauge symmetries of a given free theory. In the HS case, the common starting point is the free Fronsdal action \cite{Fronsdal:1978rb}, which schematically reads
\begin{equation}
    S^{(2)}=\frac12\int \phi_{\mu(s)} \Box\, \phi^{\mu(s)}+\ldots\,,\label{Fronsdal}
\end{equation}
with the totally symmetric doubly-traceless field $\phi_{\mu(s)}(x)$. The latter is subject to the linear gauge symmetry
\begin{equation}
    \delta_\epsilon^{(0)}\phi_{\mu(s)}=\pl_\mu\epsilon_{\mu(s-1)}\,,
\end{equation}
with a traceless gauge parameter $\epsilon$. The $...$ denote terms which depend on the choice of gauge.

It is convenient to encode tensors into polynomials in an auxiliary variable $u^\mu$ as,
\begin{equation}
\phi(x,u)=\frac1{s!}\,\phi_{\mu(s)}(x)u^{\mu(s)}\,.
\end{equation}
a formalism which we employ throughout this article. In this way the full Fronsdal action \cite{Fronsdal:1978rb} in de Donder gauge can be expressed in the simple form
\begin{equation}
S^{(2)}=\frac{s!}2\int \phi(x,\pl_u)\left(1-\frac14 u^2\pl_u^2\right)\Box\,\phi(x,u)\Big|_{u=0}\,,
\end{equation}
with canonical normalisation for the kinetic term.

Towards constructing interactions as deformations of the free theory \eqref{Fronsdal}, one assumes the existence of a non-linear action $S$ and non-linear gauge transformations $\delta_\epsilon\phi$, such that in a weak field expansion they satisfy the following boundary conditions:
\begin{subequations}
\begin{align} \label{wfact}
S&=S^{(2)}+\sum_{n>0}S^{(n+2)}\,,\\ \label{wfga}
\delta_{\epsilon}\phi_{\mu(s)}&=\delta_\epsilon^{(0)}\phi_{\mu(s)}+\sum_{n>0}\delta_\epsilon^{(n)}\phi_{\mu(s)}\,.
\end{align}
\end{subequations}
The notation $f^{(n)}$ denotes a term of order $n$ in the weak fields $\phi$.
Imposing that the non-linear action \eqref{wfact} is invariant under the non-linearly deformed gauge symmetries \eqref{wfga} then provides a framework to solve, order by order in the weak fields, for the non-linear theory. All in all, one ends up with the following system of coupled equations:
\begin{subequations}\label{Noether}
\begin{align}
\delta^{(0)}_\epsilon S^{(2)}&=0\,,\\ \label{3nc}
\delta^{(1)}_\epsilon S^{(2)}+\delta^{(0)}_\epsilon S^{(3)}&=0\,,\\ \label{4nc}
\delta^{(2)}_\epsilon S^{(2)}+\delta^{(1)}_\epsilon S^{(3)}+\delta^{(0)}_\epsilon S^{(4)}&=0\,,\\
\ldots\,,\\
\delta^{(n)}_\epsilon S^{(2)}+\sum_{k=1}^n\delta^{(n-k)}_\epsilon S^{(2+k)}+\delta^{(0)}_\epsilon S^{(2+n)}&=0\,.
\end{align}
\end{subequations}
While the first equation defines the Fronsdal theory itself, the other equations can be solved in a iterative way modulo the free equations of motion (see e.g. \cite{Berends:1984rq,Berends:1984wp,Barnich:1993vg,Joung:2012fv,Joung:2013nma}):
\begin{equation}\label{np2}
  \delta^{(0)}S^{(2+n)} \approx-\sum_{k=1}^{n-1}\delta^{(n-k)}_\epsilon S^{(2+k)}\,,
\end{equation}
and then reading off the deformations of the gauge transformations:
\begin{equation}
  \delta^{(n)}_\epsilon S^{(2)}=-\sum_{k=1}^{n-1}\delta^{(n-k)}_\epsilon S^{(2+k)}\,,
\end{equation}
The above equations will be the main subject of the following sections. We first study their formal solution up to any order $n+2$. We then focus on the  $s$-$0$-$0$-$0$ case (the prototype of Weinberg-like arguments \cite{Weinberg:1964ew}), investigating the locality properties of the solution. For the interested reader and also for pedagogical reasons we review some examples (see also section 4.1 in \cite{Taronna:2011kt}) including the Yang-Mills case in appendix~\ref{YMnoether}.

\subsection{Current exchange and Noether procedure}

A key feature of the system of equations \eqref{Noether} is that, for $n>1$, the equation for $S^{(n+2)}$ is inhomogeneous.\footnote{For $n=1$ instead, the equation is homogeneous. Covariant cubic couplings have been classified together with the corresponding deformations of the gauge symmetries in \cite{Boulanger:2008tg,Manvelyan:2010jr,Sagnotti:2010at,Fotopoulos:2010ay,Joung:2011ww,Joung:2012fv,Taronna:2012gb}, for a short review of the solution for cubic couplings and their deformations we refer to appendix \ref{cubic}.} Supposing that we have solved the Noether procedure up to a given order $n+1$, to solve at the next order $n+2$ with $n>1$ for $S^{(n+2)}$, one searches for a particular solution to the inhomogeneous equations \eqref{np2}, so that the most general solution reads:
\begin{equation}
  S^{(n+2)}=S_{h}^{(n+2)}+S_p^{(n+2)}\,.\label{mostgen}
\end{equation}
$S_h^{(n+2)}$ is an arbitrary solution to the homogeneous equation
\begin{equation}
  \delta^{(0)}S_h^{(n+2)}\approx 0\,,\label{homo}
\end{equation}
which is the equation solved by cubic couplings. $S_p^{(n+2)}$ is the particular solution to the original inhomogeneous equation \eqref{np2} which contains the information about lower order solutions.

The most general form of the homogeneous solution at order $n+2$ can be expressed as an arbitrary function $f$ of certain tensorial structures $\mathcal{K}_i$ (see \cite{Taronna:2011kt}):
\begin{equation}\label{homosol}
S_h^{(n+2)}=f^{(n+2)}({\sf s}_{ij},{\cal K}_i)\,,
\end{equation}
where ${\sf s}_{ij}=-(p_i+p_j)^2$ are the generalised Mandelstam invariants. At this point no further condition on the function $f^{(n+2)}$ arise from the Noether procedure neither at this or higher orders. Non trivial constraints on the function $f^{(n+2)}$ can arise if one imposes locality conditions.\footnote{Further non-trivial constraints arise from restricting attention to killing tensors and to the deformed symmetries generated by cubic couplings given by $\delta^{(1)}_{\bar{\epsilon}}$. These satisfy $\delta^{(0)}_{\bar{\epsilon}}\phi=0$ so that Noether procedure can be shown to be equivalent to a simple invariance condition of the homogeneous solutions $\delta^{(1)}_{\bar{\epsilon}}S_h^{(n+2)}\approx 0$. These further conditions can be in principle used to constrain homogeneous solutions dependence on Mandelstam invariants up to any order with the only obstruction being that $\delta^{(1)}_{\bar{\epsilon}}$ closes a (infinite dimensional) Lie algebra. This are related to the existence of non trivial asymptotic charges acting on the observables/invariants of the theory. We will not discuss consequences of this further condition here and refer for related discussions to section 5 of \cite{Sleight:2016xqq}.} In this case, as discussed below, the functions $f$ arising at a given order can be constrained requiring locality at the succeeding orders.

For the particular solution $S_p^{(n+2)}$, the key observation is that a solution is given by \emph{minus} the current exchange amplitude with $n+2$ external legs, built from all couplings with a number of legs strictly lower than $n+2$.\footnote{Here, by current exchange amplitude we mean the sum over exchanges in all channels.} This observation allows to formally solve the Noether procedure up to any order in perturbation theory. Because of this, the particular solution is simply obtained by classifying all non-trivial solutions \eqref{homosol} to the \emph{homogeneous} equation \eqref{homo} for each order up to $n+2$, and compute all possible exchange amplitudes involving them.

The simplest example of the above is given by the quartic case, where the exchange amplitude is given by
\begin{equation}
\mathcal{E}^{(4)}=\left[f^{(3)}(\pl_{u_1},\pl_{u_2},\pl_u)f^{(3)}(\pl_{w_3},\pl_{w_4},\pl_w)\mathcal{P}_{\sf s}(u,w)+\ldots\right]\phi_1(u_1)\phi_2(u_2)\phi_3(w_3)\phi_4(w_4)\,,
\end{equation}
with the dots denoting the contributions of ${\sf t}$ and ${\sf u}$ channels.
Above $\mathcal{P}(u,w)$ is the generating function form of the propagator numerator and $f(\pl_{u_i})$ is the generating function form of any consistent solution to the cubic Noether equations (see appendix~\ref{cubic}). It is straightforward to then show that
\begin{equation}
\delta_\epsilon^{(0)}\mathcal{E}^{(4)}=\delta_\epsilon^{(1)}S^{(3)}\,,
\end{equation}
simply using the definition of the propagator \eqref{propdef} and eqs.~\eqref{gaugedef}. This explicitly shows in full generality that the quartic Noether consistency conditions \eqref{4nc} are satisfied modulo the free equations of motion
\begin{equation}
\delta^{(0)}(-\mathcal{E}^{(4)})+\delta^{(1)}S^{(3)} \approx 0\,.
\end{equation}
Together with \eqref{mostgen} the above then provides the most general solution of the Noether procedure at quartic order, up to a solution to the homogeneous equation \eqref{homosol}.

In more physical terms, on-shell, the homogeneous solution plays the role of the S-matrix amplitude and Noether procedure simply ensures its Poincar\'e invariance which for massless particles corresponds to the condition of decoupling of the unphysical polarisations \eqref{homo} (see also the discussion in \cite{Weinberg:1964ew}).
One can generalise the above to any order, providing an alternative proof of the result obtained in \cite{Taronna:2011kt} via BRST methods. Notice, however, that Noether procedure alone at this level does not give any condition on how to relate homogeneous solutions for different $n$. What makes Noether procedure non-trivial, and its solution well defined, is a functional class space to which the full solution $S^{(n+2)}$ should be required to belong. Usually, a functional class space is a locality requirement on the actual Lagrangian couplings which can translate into appropriate convergence conditions for the derivative expansion.\footnote{The issue of locality and functional class has also recently received attention in anti-de Sitter (AdS) space  \cite{Prokushkin:1999xq,Bekaert:2014cea,Kessel:2015kna,Boulanger:2015ova,Bekaert:2015tva,Skvortsov:2015lja,Vasiliev:2016xui,Taronna:2016ats,Bekaert:2016ezc,Taronna:2016xrm,Sleight:2016hyl}. In AdS the distinction between pathological $1/\Box$-like non-localities and admissible non-localities is in fact less sharp than in flat space \cite{Taronna:2016xrm}, and the latter do arise naturally in the unfolded formalism of \cite{Vasiliev:1990en}. In particular, a seemingly well-behaved redefinition with factorially convergent coefficients multiplying the higher-derivative terms is capable of removing non-trivial interactions \cite{Taronna:2016xrm}. 

See also \cite{Rahman:2016tqc} for the study of propagation of HS fields on non-trivial backgrounds and for the relation between analyticity in curvatures and locality.}

\section{The homogeneous solution}\label{Homosol}

The previous section explained how solving the Noether procedure is equivalent to classifying the homogeneous solutions \eqref{homo} up to a given order. Once the homogeneous solution is found, its overall coefficients (the function $f$ in \eqref{homosol}) can be freely chosen, and no further condition arises from Noether procedure at \emph{any} order \emph{if no locality condition is enforced}.\footnote{See also footnote 6.} In thise case, the only non-trivial information is given by the number of independent solutions to the homogeneous equation, which in turn gives the space of free parameters arising at a given order in the weak field expansion.

The simplest and instructive example to illustrate this logic is given by the $s$-$0$-$0$-$0$ case. Using momentum space and going on-shell for the external legs gives the following most general ansatz up to total derivatives:
\begin{equation}
S^{(4)}_h=f^{(4)}(Y_{12},Y_{14},{\sf s},{\sf u})\,,
\end{equation}
in terms of a function of four variables. Above we have introduced the notation:
\begin{equation}
Y_{ij}=-i\pl_{u_i}\cdot p_j\,,
\end{equation}
together with the Mandelstam variables:
\begin{subequations}
\begin{align}
{\sf s}&=-(p_1+p_2)^2\,,\\
{\sf t}&=-(p_1+p_3)^2\,,\\
{\sf u}&=-(p_1+p_4)^2\,,\\
&{\sf s}+{\sf t}+{\sf u}=0\,.
\end{align}
\end{subequations}
The homogeneous equation \eqref{homo} then becomes the following linear differential equation for the function $f^{(4)}$:
\begin{equation}
({\sf s}\,\pl_{Y_{12}}+{\sf u}\,\pl_{Y_{14}})f^{(4)}=0\,,
\end{equation}
which admits the general solution:
\begin{equation}\label{NoetherSol}
f^{(4)}=f^{(4)}(\mathcal{K}\equiv {\sf u}\,Y_{12}-{\sf s}\,Y_{14},{\sf s},{\sf u})\,.
\end{equation}
The above fixes uniquely the tensorial structure of the quartic amplitude for given spin-s up to an arbitrary function of the Mandelstam invariants.
The following identities:
\begin{equation}
{\sf u}\,Y_{12}-{\sf s}\,Y_{14}={\sf u}\,Y_{13}-{\sf t}\,Y_{14}={\sf s}\,Y_{13}-{\sf t}\,Y_{12}\,,
\end{equation}
follow on-shell, up to total derivatives and allow to reinterpret the same tensor structure in different channels.

Similar solutions can be obtained at any order $n+2>3$ and are simply particular examples of the following curl type structures (see e.g. \cite{Joung:2013nma}):
\begin{equation}\label{Hcurl}
H_{\mu\nu}=(-ip_1)_{\mu}(\pl_{u_1})_{\nu}-(-ip_1)_{\nu}(\pl_{u_1})_{\mu}\,.
\end{equation}
Solution to the homogeneous equation can then be obtained by contracting $H_{\mu\nu}$ with two arbitrary momenta as:
\begin{equation}
\mathcal{K}_{ij}=p_i^\mu p_j^\nu H_{\mu\nu}\,,
\end{equation}
arriving at:
\begin{align}
f^{(n+2)}&=f^{(n+2)}(\mathcal{K}_{ij},{\sf s}_{ij})\,,& {\cal K}_{ij}&= {\sf s}_{1i}Y_{1j}-{\sf s}_{1j}Y_{1i}\,,
\end{align}
where we recall that ${\sf s}_{ij}=-(p_i+p_j)^2$ are the generalised Mandelstam invariants. The number of solutions is equal to the number of independent ways of selecting pairs of Mandelstam variables among the independent ones. In \cite{Taronna:2011kt} the above solution was obtained and it was also shown how they are generated within Yang-Mills theory.

An interesting feature of the curl-like structure \eqref{Hcurl} of the homogeneous solutions $S^{(n+2)}_h$ above is that they are identically gauge invariant. For this reason homogeneous solutions do not induce a deformation of the gauge transformations. At the cubic order, on the other hand, the homogeneous solution may not be of the curl-type above. This implies that the deformations of the gauge transformations at a given order are entirely a consequence of the current exchange amplitude, which is gauge invariant up to the linearised equations of motion due to the presence of cubic couplings. A corollary of this result is that any theory without cubic vertices produces deformations of the gauge transformations that can be removed by a field redefinition, and are therefore Abelian.

\section{Locality and higher spins}\label{loc}

So far we have presented the most general solution to the Noether procedure at quartic order for the simple case of one spinning external leg. We have not yet imposed any further condition on top of gauge invariance. As we have explicitly shown (and in accordance with the theorem of \cite{Barnich:1993vg}), if no extra conditions are enforced the Noether procedure admits an infinite number of $1/\Box$-type solutions up to any order in perturbation theory, parametrised by the coefficients of the solution to the homogeneous equation \eqref{homo}. However, such solutions are generically non-local, since the particular solution $S^{(4)}_p = -\mathcal{E}^{(4)}$ contains poles in the Mandelstam variables which can survive in \eqref{mostgen}. Allowing poles in the Mandelstam invariants at the level of \eqref{mostgen} trivialises Noether procedure, allowing to find solutions for any choice of the couplings. Looking for solutions \eqref{mostgen} which do not involve such poles, if any, is what makes Noether procedure non-trivial.\footnote{In AdS the situation is even more subtle as factorially convergent expansions in derivatives are sufficient to reproduce the AdS analogue of $1/\Box$ singularities \cite{Prokushkin:1999xq,Kessel:2015kna,Boulanger:2015ova,Skvortsov:2015lja,Taronna:2016ats,Taronna:2016xrm}.}

The cancellation of all $1/\Box$ singularities is a highly non-trivial condition and, in a weak field expansion, can be considered as a necessary condition for the existence of a HS theory beyond the cubic order. To anticipate the main conclusion, in this note we will see that, under some justified assumptions on the convergences of the derivative expansion and the sum over spins (which generalise the assumptions of \cite{Weinberg:1964ew}), we can show that such a cancellation cannot take place for massless HS theories in flat space.

In order to investigate this problem more closely, a key step is to study the structure of the particular solution to the Noether procedure, namely \emph{minus} the current exchange amplitude. This indeed involves the $1/\Box$ non-localities which, as we have argued, must cancel when considering the full solution \eqref{mostgen}.

In the following, for ease of notation and to keep the discussion as simple as possible, we shall focus on the $s$-$0$-$0$-$0$ case (the prototype of Weinberg type arguments). We will discuss more general cases in appendix \ref{s1s1}. The cubic couplings which enter the exchange diagrams are given by $s_1$-$s_2$-$0$ couplings (see appendix \ref{cubic}):\footnote{Above we have introduced arbitrary coefficients instead of referring to exponential generating functions, as done in \cite{Taronna:2011kt}. In this way the discussion is more transparent. Notice, however, that the discussion carried out below has the same level of generality as in \cite{Taronna:2011kt} where it was assumed that the generating functions were defined up to arbitrary coefficients for the various independent structures.}
\begin{equation}
S^{(3)}=\sum_{s_1,s_2}g_{s_1,s_2,0}\,Y_1^{s_1}\,Y_2^{s_2}\phi_1\phi_2\phi_3\,,\label{s1s20}
\end{equation}
where the coefficients $g_{s_1,s_2,0}$, for now, are left arbitrary. 
 In the following we will also assume, without loss of generality, that the coupling constants $g_{s_1,s_2,s_3}$ are cyclic $g_{s_1,s_2,s_3}=g_{s_2,s_3,s_1}=g_{s_3,s_1,s_2}$ to have manifest Bose symmetry between different legs.

\subsection{The propagator and the exchange}

At the four-point level the exchange can be constructed in terms of three planar structures, which can be dressed with appropriate Chan-Paton factors. Denoting $\mathcal{E}_{1234}$ the basic planar structure with poles in the ${\sf s}$ and ${\sf u}$ channels, one then recovers the full exchange by summing over inequivalent permutations as
\begin{equation}
\mathcal{E}^{(4)}=\sum_\sigma \mathcal{E}_{1\sigma(2)\sigma(3)\sigma(4)}\text{Tr}(\phi_1 \phi_{\sigma(2)} \phi_{\sigma(3)} \phi_{\sigma(4)})\,.
\end{equation}
While the full exchange needs to be considered in general, it is sometimes convenient to restrict to a given planar structure, say $\mathcal{E}_{1234}$. Indeed, requiring that the amplitude can be dressed with Chan-Paton factors implies that Noether procedure equations \eqref{Noether} should hold separately for each color-ordered term. In the latter case one solves the Noether procedure under the assumption that the fields and their interactions can be colored with $O(n)/U(n)$ Chan-Paton factors.\footnote{This situation is reminiscent of open string theory \cite{Paton:1969je,Schwarz:1982jn}, which can also be generalised to the HS case as discussed in \cite{Metsaev:1991nb}. An interesting problem is in this respect the fate of the graviton when considering the tensionless limit, as discussed in the introduction of \cite{Francia:2007qt}.} The HS fields transform under the symmetric or anti-symmetric representation depending on the spin (even or odd). Notice that the singlet sector for even spins requires some dedicated study as all color orderings will contribute to the same Noether equation and some of the obstructions may compensate each other. 

The exchange amplitude for external scalars was computed in \cite{Bekaert:2009ud}, and presented in terms of the propagator obtained in \cite{Francia:2007qt} (see \cite{Taronna:2010qq,Sagnotti:2010at} for further details). The flat space propagator in the de-Donder gauge solves the equation:
\begin{equation}\label{propdef}
\left(1-\frac14 u^2\pl_u^2\right)\Box\,\mathcal{P}(x,u;y,v)=-\delta(x-y)[(u\cdot v)^s+\ldots]\,,
\end{equation}
where the $\ldots$ ensure the doubly-tracelessness projection $(\pl_u^2)^2[(u\cdot v)^s+\ldots]=0$.
It can be conveniently encoded, up to gradient terms, in a Gegenbauer polynomial 
\begin{equation}
\mathcal{P}_r(u,v)=-\frac1{2^r\, r!}\frac{i}{p^2}\,\frac1{(\tfrac{d}2-2)_r}\,(u^2 v^2)^{\tfrac{r}2}C_r^{d/2-2}\,\left(\frac{u\cdot v}{\sqrt{u^2 v^2}}\right)\,.
\end{equation}
For convenience we used canonical normalisation for the kinetic term and introduced the ascending pochhammer symbol $(x)_n=\Gamma(x+n)/\Gamma(x)$. Above, index contraction is performed via the transverse space metric:
\begin{equation}
\Pi_{\mu\nu}=\eta_{\mu\nu}-p_\mu \bar{p}_\nu-p_\nu \bar{p}_\mu\,,
\end{equation}
with $p_\mu$ the exchanged momentum and $\bar{p}_\mu$ an auxiliary null vector with the property $p\cdot \bar{p}=1$ (see \cite{Francia:2007qt}). In the flat space case the Fronsdal currents coming from the consistent cubic couplings admit an improvement to a conserved currents, which can be explicitly constructed (see appendix~\ref{cubic}). One can then drop all dependence on $p_\mu$ and $\bar{p}_\mu$.

In the following it is convenient to restrict to the part of the exchange which is non-local and is proportional to $1/\Box$. This part is the pathological non-local contribution to the solution \eqref{mostgen}, which we need to compensate in order to define a proper functional class space and thus a non-trivial solution to the Noether procedure. It is important to stress that focusing on the non-local part of the exchange implicitly assumes certain convergence properties for the sum over spins when all contact terms coming from different exchanges are summed together, which we detail in section~\ref{convergence}. It might be indeed possible that, upon summation over spins, the infinite number contact contributions (with each being polynomial in the Mandelstam variables) generate poles. However, as discussed in section \ref{convergence}, the contact part of the exchange is convergent. Furthermore this assumption is justified in a field theory context, where it is expected that all singularities of the amplitude originate from propagators and do not come from infinite sums of contact terms. This assumption is also used in Weinberg's argument \cite{Weinberg:1964ew}, in order to drop the contact terms in the soft limit of the amplitude. In section \ref{convergence} we also discuss the possibility of including additional improvement terms and their potential role in this game, as improvements can easily be used to generate non-convergent contributions to the current exchange. 

The result for the planar exchange of a spin-$r$ particle where $\phi_1$ is the only spinning polarisation of spin $s_1$ is given by
\begin{equation}
\mathcal{E}^{[r]}_{1234}=-\frac1{{\sf s}}\,g_{s_1,0,r}\,g_{r,0,0}\,Y_{12}^{s_1}\left(\frac{{\sf t}-{\sf u}}4\right)^r-\frac1{{\sf u}}\,g_{s_1,r,0}\,g_{r,0,0}\,(-Y_{14})^{s_1}\left(\frac{{\sf t}-{\sf s}}4\right)^r+\text{local}\,.\label{planar}
\end{equation}
Above the subscript $_{1234}$ stands for the fact that the above part of the full exchange comes with the Chan-Paton trace $\text{Tr}(\phi_1\phi_2\phi_3\phi_4)$.
The full non-planar exchange relevant for the color-singlet sector is given instead by the following three contributions:
\begin{multline}
\mathcal{E}^{[r]}=-\frac1{{\sf s}}\,g_{s_1,0,r}\,g_{r,0,0}\,Y_{12}^{s_1}\left(\frac{{\sf t}-{\sf u}}4\right)^r-\frac1{{\sf u}}\,g_{s_1,r,0}\,g_{r,0,0}\,(-Y_{14})^{s_1}\left(\frac{{\sf t}-{\sf s}}4\right)^r\\-\frac1{{\sf t}}\,g_{s_1,r,0}\,g_{r,0,0}\,(-Y_{13})^{s_1}\left(\frac{{\sf s}-{\sf u}}4\right)^r+\text{local}\,.\label{nonplanar}
\end{multline}
Above, $r$ labels the spin of the exchanged particle to avoid confusion with the Mandelstam invariant ${\sf s}$. For ease of notation we have not displayed the form of the part of the exchange that is a polynomial function of the Mandelstam invariants (i.e. the local part), which we discuss in section~\ref{convergence}.

\subsection{Looking for a local solution}\label{lookingfor}

So far we have gathered all the ingredients required to study the locality properties of the most general quartic solution with a single HS external leg in flat space. Focusing on a single color ordered contribution and fixing the external spin to be $s_1$, the quartic solution reads:
\begin{multline}\label{quarticsol}
{S}^{(4)}_{1234}=\frac{1}{2^{s_1}}f^{(4)}_{s_1}({\sf s},{\sf u})({\sf u} Y_{12}-{\sf s} Y_{14})^{s_1}\\+\sum_r\left[\frac1{{\sf s}}\,g_{s_1,0,r}\,g_{r,0,0}\,Y_{12}^{s_1}\left(\frac{{\sf t}-{\sf u}}4\right)^r+\frac1{{\sf u}}\,g_{s_1,r,0}\,g_{r,0,0}\,(-Y_{14})^{s_1}\left(\frac{{\sf t}-{\sf s}}4\right)^r+\text{local}\right]\,.
\end{multline}
As one can explicitly see, there are infinitely many coefficients to be fixed in the function $f$ and, as well, infinitely many arbitrary cubic coupling constants in the current exchange. The above is the most general non-local solution to Noether procedure found in \cite{Taronna:2011kt}. The cubic couplings so far have been left unfixed. In particular, the latter are not fixed at cubic order in the Noether procedure, but in principle can be determined by higher-order consistency conditions. On the other hand, as we have already emphasised, it is crucial that the couplings conspire to cancel \emph{all} $1/\Box$ non-localities present in \eqref{quarticsol}.\footnote{Indeed, a particularly simple and possibly trivial solution among the ones above is for instance
\begin{equation}
f_{s_1}^{(4)}({\sf s},{\sf u})\equiv 0\,,\label{trivialS}
\end{equation}
where all $1/\Box$ non-localities are kept.
With this choice of quartic interactions \emph{all} $4$-pt S-matrix amplitudes vanish identically, regardless of the choice of cubic coefficients $g_{s_1,s_2,s_3}$. We thus end up with a theory with vanishing S-matrix amplitudes, consistent with HS symmetry \cite{Sleight:2016xqq}. It is however unclear how to fix the cubic coupling constant in this case, as the above solution with trivial S-matrix is available for any choice of $g_{s_1,s_2,s_3}$.}

Putting aside the above simple solutions, in the following we investigate whether or not the above $1/\Box$ non localities can be cancelled, leading to a local quartic coupling. Notice, however, that locality cannot constrain the part of the function $f$ which is an entire function of the Mandelstam invariants -- at least beyond fixing the asymptotic behaviour for their overall coefficients. One thus sees a generic feature of Noether procedure: at any order one finds infinitely many local homogeneous solutions whose overall coefficient cannot be fixed at the same order. A similar feature is of course present at cubic order with the only difference being that, due to the lack of Mandelstam invariants, one only has a finite number of non-trivial structures. The free local structures which one can introduce at any given order may be fixed only at the \emph{next} or higher orders.

Since the entire part of the function $f$ cannot be fixed by consistency at this order, without loss of generality it is convenient to restrict to only the pole part of ${S}^{(4)}$, which as at most single poles in the Mandelstam invariants.
One can thus consider an expansion for the function $f({\sf s},{\sf u})$ of the type:\footnote{We emphasise that the assumption to drop higher-order poles in any of the Mandelstam variables plays here a key role. If we were to forgo this assumption it is relatively straightforward to compensate all single poles in the $\sf s$-channel with the choice:
\begin{equation}\label{fs1bis}
f_{s_1}({\sf s},{\sf u})=\frac1{{\sf su}^{s_1}}\sum_{n=0}^\infty g_{s_1000}^{[n]}\left(\frac{{\sf t}-{\sf u}}{4}\right)^n\,.
\end{equation}
Combining this with the tensor structure \eqref{quarticsol} we then get:
\begin{equation}
S^{(4)}_{s}=\frac{Y_{12}^{s_1}}{{\sf s}}\left[\sum_{n\geq 0} g_{s_1000}^{[n]}\left(\frac{{\sf t}-{\sf u}}{4}\right)^{n} +\sum_{r\geq 0} g_{s_1,0,r}\,g_{r,0,0}\left(\frac{{\sf t}-{\sf u}}{4}\right)^{r}\right]\,.
\end{equation}
Therefore no problem arise looking at a single channel for a given quartic coupling.
However the ${\sf u}$-channel will now develop higher-order poles, as can be seen from:
\begin{align}
\frac1{{\sf s}}\frac1{{\sf u}^{s_1}}({\sf u} Y_{12}-{\sf s} Y_{14})^{s_1}=\frac1{{\sf s}}\,Y_{12}^{s_1}+\sum_{k=1}^{s_1} \binom{s_1}{k}\frac{{\sf s}^{k-1}}{{\sf u}^k}Y_{12}^{s_1-k}\,(-Y_{14})^{k}\,.
\end{align}
It is also important to stress that such higher-order poles will contribute to the quartic contact term and can only be seen when decomposing the same contact term in the ${\sf u}$-channel, making it non-local even in the case that all single poles in the ${\sf s}$-channel cancel.
}
\begin{equation}\label{fs1}
f_{s_1}({\sf s},{\sf u})=\frac2{{\sf su}}\sum_{n=0}^\infty g_{s_1000}^{[n]}\left(-\frac{{\sf s}+{\sf u}}{2}\right)^n\,,
\end{equation}
with single poles only in ${\sf s}$ and ${\sf u}$. This leads to:
{\allowdisplaybreaks
\begin{align}
S^{(4)\,n.l}_{1234}=\frac{Y_{12}^{s_1}}{{\sf s}}&\left[(-1)^{s_1-1}\,\sum_{n\geq 0} g_{s_1000}^{[n]}\left(\frac{{\sf t}-{\sf u}}{4}\right)^{n+s_1-1} +\sum_{r\geq 0} g_{s_1,0,r}\,g_{r,0,0}\left(\frac{{\sf t}-{\sf u}}{4}\right)^{r}\right]+\nonumber\\
+\frac{(-Y_{14})^{s_1}}{{\sf u}}&\left[(-1)^{s_1-1}\,\sum_{n\geq 0} g_{s_1000}^{[n]}\left(\frac{{\sf t}-{\sf u}}{4}\right)^{n+s_1-1} +\sum_{r\geq 0} g_{s_1,r,0}\,g_{r,0,0}\left(\frac{{\sf t}-{\sf u}}{4}\right)^{r}\right]\,.
\end{align}}
Above we have extracted the non-local part of the sum of the homogeneous solution and particular solution \eqref{quarticsol}. Assuming now the existence of a domain in which the above sums converge uniformly, one can then gather the two sums into a single sum adding up terms of the same degree in $({\sf t}-{\sf u})$ or $({\sf t}-{\sf s})$, arriving to
\begin{align}\label{dangerous}
S^{(4)\,n.l}_{1234}=\frac{Y_{12}^{s_1}}{{\sf s}}&\left[(-1)^{s_1-1}\,\sum_{n\geq 0} \left[g^{[n]}_{s_1000}+(-1)^{s_1-1}g_{s_1,0,n+s_1-1}\,g_{n+s_1-1,0,0}\right]\left(\frac{{\sf t}-{\sf u}}{4}\right)^{n+s_1-1} \right.\nonumber\\&\left.\hspace{200pt}+\sum_{r= 0}^{s_1-2} g_{s_1,0,r}\,g_{r,0,0}\left(\frac{{\sf t}-{\sf u}}{4}\right)^{r}\right]\nonumber\\
+\frac{(-Y_{14})^{s_1}}{{\sf u}}&\left[(-1)^{s_1-1}\,\sum_{n\geq 0} \left[g^{[n]}_{s_1000}+(-1)^{s_1-1}g_{s_1,n+s_1-1,0}\,g_{n+s_1-1,0,0}\right]\left(\frac{{\sf t}-{\sf s}}{4}\right)^{n+s_1-1} \right.\nonumber\\&\left.\hspace{200pt}+\sum_{r=0}^{s_1-2} g_{s_1,r,0}\,g_{r,0,0}\left(\frac{{\sf t}-{\sf s}}{4}\right)^{r}\right]\,.
\end{align}
The above equation summarises one of the main results of \cite{Taronna:2011kt}. Indeed while it is possible to fix: 
\begin{equation}
g^{[n]}_{s_1000}=(-1)^{s_1}g_{s_1,0,n+s_1-1}\,g_{n+s_1-1,0,0}\,,
\end{equation}
to cancel the $1/\Box$ non-locality for infinitely many exchanges with spin $s\geq s_1-1$, it is not possible to cancel the non-locality associated to exchanged spins with $r\leq s_1-2$. One also recovers that there is no problem for $s_1=0,1$, while the first obstruction arises for $s_1=2$ and is proportional to $g_{s_1,0,0}\,g_{0,0,0}$. The reason we see a problem already for $s_1=2$ is related to the fact that we have restricted the attention to a single planar contribution to the exchange with a Chan-Paton factor of the type $\text{Tr}(\phi_1\phi_2\phi_3\phi_4)$ and $\phi_i=\phi_a T_i^a$. The above inconsistency implies that one cannot find a local $2$-$0$-$0$-$0$ coupling for a colored spin-2 field interacting with scalars,\footnote{More obstructions would follow for colored spin-2 by considering more external spinning legs.} in accordance with known no-go theorems in flat space \cite{Boulanger:2000rq}. In order to analyse the case of gravitational interactions it is needed to consider the singlet sector for even spins (associated to $O(1)$ Chan-Paton factors \cite{Schwarz:1982jn}) and to this end one must study the full exchange amplitude, since all color orderings contribute to the singlet. In this case one obtains
\begin{multline}
\mathcal{S}^{(4)}=\frac{1}{2^{s_1}}f({\sf s},{\sf u})({\sf u} Y_{12}-{\sf s} Y_{14})^{s_1}+\sum_r\left[\frac1{{\sf s}}\,g_{s_1,0,r}\,g_{r,0,0}\,Y_{12}^{s_1}\left(\frac{{\sf u}-{\sf t}}4\right)^r\right.\nonumber\\
\left.+\frac1{{\sf u}}\,g_{s_1,r,0}\,g_{r,0,0}\,(-Y_{14})^{s_1}\left(\frac{{\sf s}-{\sf t}}4\right)^r+\frac1{{\sf t}}\,g_{s_1,r,0}\,g_{r,0,0}\,(-Y_{13})^{s_1}\left(\frac{{\sf u}-{\sf s}}4\right)^r+\text{local}\right]\,.
\end{multline}
As before, the analysis of locality properties can be done restricting the attention to the part of the above general solution proportional to $1/\Box$. 
Since the singlet sector is only meaningful for even spins, it is convenient to rewrite the most general homogeneous solution in an on-shell equivalent form given by:
\begin{equation}
{S}^{(4)}_{h}=f_{s_1}({\sf s},{\sf t},{\sf u})\left({\sf tu}\, Y_{12}^2+{\sf us}\, Y_{13}^2+{\sf st}\, Y_{14}^2\right)^{s_1/2}\,.
\end{equation}
One can then write down the most general non-local homogeneous solution with at most single poles in the Mandelstam invariants as:\footnote{The dependence on the Mandelstam variables is fixed up to local terms, which we can drop at this level as before.}
\begin{equation}
f_{s_1}({\sf s},{\sf t},{\sf u})=\frac1{2^{s_1}}\,\frac1{{\sf stu}}\sum_{n\geq 0}g^{[n]}_{s_1,0,0,0}\left(\frac{{\sf st}+{\sf tu}+{\sf us}}{4}\right)^{n/2}\,.
\end{equation}
The non-local part of the most general solution for ${S}^{(4)}$ then reads:
\begin{align}
S^{(4)}_{n.l.}=\frac{Y_{12}^{s_1}}{{\sf s}}&\left[\sum_{n\geq 0} g^{[n]}_{s_1000}\,(-1)^{\tfrac{s_1+n}2-1}\left(\frac{{\sf t}-{\sf u}}{4}\right)^{n+s_1-2} +\sum_{r\geq 0} g_{s_1,0,r}\,g_{r,0,0}\left(\frac{{\sf t}-{\sf u}}{4}\right)^{r}\right]\nonumber\\
+\frac{Y_{13}^{s_1}}{{\sf t}}&\left[\sum_{n\geq 0}g^{[n]}_{s_1000}\,(-1)^{\tfrac{s_1+n}2-1}\left(\frac{{\sf s}-{\sf u}}{4}\right)^{n+s_1-2} +\sum_{r\geq 0} g_{s_1,r,0}\,g_{r,0,0}\left(\frac{{\sf t}-{\sf u}}{4}\right)^{r}\right]\nonumber\\
+\frac{Y_{14}^{s_1}}{{\sf u}}&\left[\sum_{n\geq 0}g^{[n]}_{s_1000}\,(-1)^{\tfrac{s_1+n}2-1}\left(\frac{{\sf t}-{\sf s}}{4}\right)^{n+s_1-2} +\sum_{r\geq 0} g_{s_1,r,0}\,g_{r,0,0}\left(\frac{{\sf t}-{\sf u}}{4}\right)^{r}\right]\,.
\end{align}
Assuming the existence of a domain in which the above series in $n$ and $r$ converge uniformly allows to combine the two sums as:
\begin{align}
S^{(4)}_{n.l.}=\frac{Y_{12}^{s_1}}{{\sf s}}&\left[\sum_{n\geq 0} \left[g^{[n]}_{s_1000}\,(-1)^{\tfrac{s_1+n}2-1}+ g_{s_1,0,n+s_1-2}\,g_{n+s_1-2,0,0}\right]\left(\frac{{\sf t}-{\sf u}}{4}\right)^{n+s_1-2} \right.\\&\hspace{200pt}\left.+\sum_{r=0}^{s_1-4} g_{s_1,0,r}\,g_{r,0,0}\left(\frac{{\sf t}-{\sf u}}{4}\right)^{r}\right]+\ldots\,,\nonumber
\end{align}
where the $\ldots$ stand for the contributions of the other channels which have a similar structure. As in the planar case one can now impose:
\begin{equation}
g^{[n]}_{s_1000}=(-1)^{\tfrac{s_1+n}2}\,g_{s_1,0,n+s_1-2}\,g_{n+s_1-2,0,0}\,,
\end{equation}
to cancel most of the non-local part of the quartic vertex. However, a leftover non-local obstruction is present for $s_1\geq 4$ and for $r\leq s_1-4$ and is proportional to $g_{s_1,0,r}\,g_{r,0,0}\neq 0$. Remarkably, if $s_1=4$ the only obstruction is proportional to $g_{4,0,0}g_{0,0,0}$. Therefore, since in the Metsaev theory \cite{Metsaev:1991nb,Metsaev:1991mt} $g_{0,0,0}\equiv 0$, the obstruction has vanishing coefficient. However, the obstruction cannot be avoided for $s_1>4$ (or in $d>4$ where $g_{0,0,0}$ is not expected to vanish for the type-A HS theory). In this case the $r=2$ contribution produces a $1/\Box$ obstruction in the contact term, which cannot be cancelled.\footnote{During the MIAPP workshop in Munich May 2016, I became aware of the result by R.Roiban and A.Tseytlin \cite{TseytlinMIAPP} that such a local $4$-$0$-$0$-$0$ solution exists through explicit computation in Metsaev theory which rely on $g_{0,0,0}=0$. This is in accordance with the result of \cite{Taronna:2011kt} eq. (5.12) and with the result presented here. Notice that such a local solution is only possible in the singlet sector and does not exist when non-trivial Chan-Paton factors are included. We would like to stress here that the existence of such a local solution is related to the vanishing of $g_{0,0,0}$ and remains true for any choice of the leftover couplings, beyond the result \cite{TseytlinMIAPP,Roiban:2017iqg}. This is also in agreement with Weinberg theorem, since in the soft limit, the offending pole arises in this case only for a scalar exchange as discussed in \cite{Taronna:2011kt}. In general, the leading contribution in the soft limit is proportional to $g_{s,r,r}$ with $r<s$ and the spin $r$ particle propagating inside the exchange -- see \cite{Taronna:2011kt}, section (4.4).} The above discussion also shows explicitly that there is no problem for gravitational couplings of the type $2$-$0$-$0$-$0$, as soon as we are in the singlet sector. This is due to the compensation between all channels -- namely, colored gravity would not pass the above non-locality test in flat space but Einstein gravity is recovered as expected.

To summarise, above we have fixed part of the quartic couplings in the HS theory in terms of the cubic couplings by requiring maximal cancellation of the $1/\Box$ non-localities. This requirement was referred to in \cite{Taronna:2011kt} as the \emph{minimal scheme}. Let us however stress that focusing only on $s$-$0$-$0$-$0$ amplitudes does not give a sufficient number of constraints to fix cubic coupling coefficients, while our conclusions and the nature of the obstruction do not depend on the value of the cubic couplings as soon as $g_{s_1,r,0}\,g_{r,0,0}\neq 0$ for $r\leq s_1-4$. As we show in appendix \ref{s1s1}, considering more complicated quartic couplings and the associated maximal cancellation of $1/\Box$ non-localities, puts further non-trivial constraints on the cubic coupling coefficients (however some $1/\Box$ non-localities still remain). We also show in appendix \ref{singleC} that for any choice of the external spins only a finite number of obstructions is present. This means that the problem arises from low spin exchanges rather than higher spin exchanges, and that the infinite remainder of the sum over spins does not give any locality obstruction of this type.

To conclude this section, it might be interesting to investigate the solution we have just found for the case of Metsaev's cubic couplings \cite{Metsaev:1991nb,Metsaev:1991mt}:\footnote{Notice that Metsaev's solution does not admit a local covariant form for all of its cubic couplings including the $s_1$-$s_2$-$0$ case, and so we will not attempt to compute exchange amplitude in Metsaev's theory using the covariant formalism considered here but just limit ourselves to fix $g_{s_1,s_2,s_3}$ as in Metsaev's theory. For the discussion of the exchange amplitude in the light-cone language we refer to section \ref{lightcone}.}
\begin{equation}
g_{s_1,s_2,s_3}=\frac1{\Gamma(s_1+s_2+s_3)}\,.
\end{equation}
In this case, dropping the non-local terms, the singlet sector solution reads:
\begin{equation}
g_{s_1000}^{[n]}=(-1)^{\tfrac{s_1+n}2}\frac1{\Gamma(2s_1+n-2)\Gamma(n+s_1-2)}\,,
\end{equation}
together with the planar solution
\begin{equation}
g_{s_1000}^{[n]}=(-1)^{s_1}\frac1{\Gamma(2s_1+n-1)\Gamma(n+s_1-1)}\,.
\end{equation}
The two solutions coincide (up to local terms) for generic even HS exchanges upon summing over all planar contributions in the singlet sector. One can also easily re-sum the above series in terms of Bessel or Hypergeometric functions, but since it is not needed for the following discussion we will not give the corresponding results here.

\subsection{Some caveats}\label{convergence}

Before concluding, it may be useful for the reader to discuss further the assumptions of the analysis carried out in this note. The key assumptions we have considered are unitarity and the existence of a region of uniform convergence for the sums over derivatives and spins in all channels. These assumptions ensure that all poles and singularities of the exchange cannot be generated in any channel from infinite sums of contact terms. A similar assumption is also natural in the context of S-matrix theory, where on-shell cubic vertices account for the full singularity of the amplitudes (see e.g. \cite{Weinberg:1964ew}). 

One may wonder if it is conceivable to drop such assumptions and look for homogeneous solutions which do not admit (simultaneous) expansion in the ${\sf s}$ and ${\sf u}$ channels. In this case, the series expansion \eqref{fs1} will be only conditionally convergent for ${\sf s}\to 0$ and its ${\sf u}\to 0$ expansion should be defined up to analytic continuation, if at all. The implications of this scenario for the locality of the solution $S^{(4)}$ is unclear, as it is natural to assume that an admissible contact term, even when unbounded in derivatives, should have a series expansion which is uniformly convergent in all channels. Failure of uniform convergence is indeed related to appearance of singularities in the contact term, which are the signal of non-localities. In this type of scenario it is also to be expected that the analytic continuation may lead to consider solutions where the function $f({\sf u},{\sf s})$ is a distribution (see e.g. \cite{Joung:2015eny,Beccaria:2016syk,Sleight:2016xqq} for discussions in this direction).

\subsubsection*{Improvements}

We emphasise that, in our opinion, a key problem in dropping the uniform convergence assumption is the possibility of allowing improvement terms which, even if local at fixed spins, give a singular contribution to the exchange upon summing over spins. Such non-convergent contact terms can be generated from local improvements at cubic order and would allow to cancel physical singularities upon considering the analytic continuation of the sum over spins. Therefore, they should be considered non-admissible from a field theory perspective (see e.g. section $3.3$ of \cite{Taronna:2016ats}).
More in detail, to each cubic coupling \eqref{cub} one can add a local improvement as:
\begin{equation}
g_{s_1,s_2,s_3}^{[k]}f_{s_1,s_2,s_3}^{[k]}\rightarrow g_{s_1,s_2,s_3}^{[k]}f_{s_1,s_2,s_3}^{[k]}+\Delta f_{s_1,s_2,s_3}^{[k]}\,.
\end{equation}
so that choosing
\begin{equation}
    \Delta f_{s_1,s_2,s_3}^{[k]}=c_{s_1,s_2,s_3}^{[k]}f_{s_1,s_2,s_3}^{[k]}\Box_3\,,
\end{equation}
with $c_{s_1,s_2,r}\sim O(1)$ for $r\to\infty$ one generates non-convergent contact terms. In the case of one spinning leg discussed in this note, for example, such improvement would generate a contact term proportional to $g_{s_1,0,r}c_{r,0,0}$ and with the same structure as the original full exchange \eqref{exchfull2}, but multiplied by an extra Mandelstam variable ${\sf s}$ so as to cancel the pole and produce a local contact term for each $r$. Therefore, allowing such a redefinitions can yield further pole contributions from the sum over spins of the improvement terms. These can indeed re-sum to a single pole of the type $\frac1{{\sf s}+i\epsilon}$. Usually these spurious poles are required to cancel as they would imply propagation of additional excitations (see below for more details). In the HS context we cannot exclude the possibility that they could play a role to compensate the non-local obstruction discussed under the assumption of uniform convergence. Notice, however, that such improvements should be considered non-admissible, for tuning appropriately the coefficients $c_{s_1,s_2,s_3}^{[k]}$ would allow to even cancel or modify the existing poles in the exchange amplitude. In this case, using similar arguments as in \cite{Taronna:2016ats}, one would call admissible only those improvements which do not introduce spurious singularities upon summing over spins. In this sense, dropping the assumption of uniform convergence seems to be beyond the framework of (pseudo-)local field theories.

\subsubsection*{Uniform convergence of contact terms}

Owing to the above discussion about uniform convergence, in this section we also take the opportunity to discuss a bit more closely the contact terms arising in the current exchange \eqref{planar} and \eqref{nonplanar}. We show that the sum over spins in this case is indeed uniformly convergent. These contributions are, however, not universal. As discussed above, they depend on the field frame chosen at cubic order and in particular on the choice of improvement terms $\Delta f_{s_1,s_2,s_3}^{[k]}$. In particular, adding improvements produces such contributions at the level of the exchange. In the following we compute the full exchange amplitude for the couplings $f_{s_1,s_2,s_3}^{[k]}$.

For ease of notation and without loss of generality it is convenient to restrict our attention to the $s$-$0$-$0$-$0$ case, where keeping all contact terms one arrives at
\begin{equation}\label{exchfull2}
\mathcal{E}^{[r]}_{\sf s}=\frac1{{\sf s}}\,Y_{1}^{s_1}\,\sum_{k=0}^{[r/2]}\alpha_k^{[r]}g_{s_1,0,r}g_{r,0,0}\left(\frac{{\sf t}-{\sf u}}{4}\right)^{r-2k}\left(\frac{{\sf s}}{2}\right)^{2k}\,,
\end{equation}
where the coefficients come from the expansion of the Gegenbauer polynomial:
\begin{equation}
\alpha_k^{[r]}=\frac{(-1)^k}{2^r}\frac{(\tfrac{d}{2}-2)_{r-k}}{(\tfrac{d}{2}-2)_{r}}\frac{r!}{k!(r-2k)!}\,.
\end{equation}
One can then study the convergence of the sum over $r$. The above is uniformly convergent whenever $g_{s_1,0,r}\,g_{r,0,0}\sim \frac1{(r!)^2}$ for $r\to\infty$, which is for instance verified by the flat limit of the holographically reconstructed couplings and Metsaev theory, justifying our assumptions. Similar conclusions can be drawn from the scalar analysis of \cite{Bekaert:2009ud}. 

\subsubsection*{Additional excitations}

Finally, another available possibility is the reinterpretation of non-local vertices as the signal of the presence of additional excitations (see e.g. \cite{Henneaux:2013vca} where this possibility was also considered in a similar context). All non-local vertices we have found have indeed the structure of a current exchange amplitude:
\begin{multline}
    S^{(4)}_{n.l.}\sim \left(\frac{Y_{12}^{s_1}}{{\sf s}}\sum_{r=0}^{s_1-4} g_{s_1,0,r}\,g_{r,0,0}\left(\frac{{\sf t}-{\sf u}}{4}\right)^{r}+\text{local}\right)+\text{crossed channels}\\\equiv -\sum_{r=0}^{s_1-4}\mathcal{E}^{[r]}_{\sf s}+\text{crossed channels}\,.
\end{multline}
It is therefore \emph{always} possible in principle to attribute such non-local contributions to exchanges coming from additional excitations.\footnote{It would be interesting to analyse this possibility in the light-cone gauge where the only known unitary solutions are of the form of color dressings of the minimal theory.} Introducing excitations $\bar{\phi}$ of spins from $0$ to infinity with couplings of the type $\bar{g}_{s_1,s_2,\bar{s}_3}\phi_1 \phi_2\bar{\phi}_3$, the exchange of $\bar{\phi}$ would then generate the following contribution:\footnote{Notice that such spectrum is formally the same as considered by Metsaev in \cite{Metsaev:1991mt} and one may argue that Metsaev's solution is unique if one assumes unitarity and that the extra fields couple minimally to gravity.}
\begin{equation}
    \mathcal{E}^{[r]}_{\sf s}\sim -\gamma\,\frac{Y_{12}^{s_1}}{{\sf s}}\sum_{r=0}^{\infty} \bar{g}_{s_1,0,\bar{r}}\,\bar{g}_{\bar{r},0,0}\left(\frac{{\sf t}-{\sf u}}{4}\right)^{r}+\text{crossed channels}
\end{equation}
with $\gamma$ the kinetic term sign for the new excitation, which is positive in the unitary case. For simplicity above we have considered only those amplitudes where $\bar{\phi}$ propagates internally. Our analysis then requires a non vanishing $\bar{g}_{\bar{s},0,0}$ for any $\bar{s}>0$ (recall that $g_{0,0,0}=0$ in 4d). In the light of the results presented in section \ref{lookingfor}, this then implies that the $\bar{\phi}$ quartic interactions $\bar{\phi}\phi\phi\phi$ are also subject to the non-local obstruction mentioned in section \ref{lookingfor} (see also appendix~\ref{s1s1}). To compensate these further non-localities a \emph{further} infinite set of fields would be required. It is therefore not clear if this procedure can terminate.\footnote{A toy model where to investigate this possibility may be the case of colored gravity where similar non-local obstruction are present.}

Let us also comment on the S-matrix interpretation of our result. As we have stressed, studying the most general homogeneous solution for the purpose of solving Noether procedure is equivalent to a classification of possible S-matrix amplitudes for massless HS fields. The result \eqref{NoetherSol}, and the corresponding non-local obstruction, can be summarised by the statement that there is no S-matrix structure with single poles in all channels and which at the same time factorises on low spin exchanges $r\leq s-4$ ($s$ is the spin of the external excitation). From an S-matrix perspective, factorisation into lower point amplitudes is a necessary condition for unitarity. We can then argue that whenever the dangerous lower-spin exchanges are non-vanishing the corresponding S-matrix amplitude is not unitary, as the only way to cancel the latter exchanges is to violate the condition of positivity for the partial wave expansion of the S-matrix amplitude. Notice that the existence of a S-matrix structure factorising on a given exchange is a necessary condition for the theory to be unitary or local and is independent on the spectrum of the theory. This underlines the closed ties between unitarity at the S-matrix level and locality in field theory.

Another possible way out is to go to the light-cone gauge, which we explore in the next section.

\section{Light-cone}\label{lightcone}

In view of the recent revival of higher-spins in the light-cone/spinor-helicity formalism \cite{Bengtsson:2016jfk,Bengtsson:2016alt,Conde:2016izb,Bengtsson:2016hss,Sleight:2016xqq,Ponomarev:2016lrm,Haehnel:2016mlb,Adamo:2016ple}, in this section we consider the light-cone reduction of the solution \eqref{NoetherSol} to the Noether procedure described in this note. At cubic order, in comparison to the covariant formulation, the light-cone gauge permits more structures compatible with Poincar\'e invariance \cite{Bengtsson:1983pd,Bengtsson:1986kh}, which have been shown to be crucial for quartic consistency \cite{Metsaev:1991mt,Metsaev:1991nb}. It is therefore interesting to consider the quartic solutions discussed in this note in the light-cone gauge.

Going to the light-cone amounts to:
\begin{align}
x^\pm&\equiv \frac{x^0\pm x^3}{\sqrt{2}}\,,& z&=\frac{x^1+i x^2}{\sqrt{2}}\,,& \bar{z}&=\frac{x^1-i x^2}{\sqrt{2}}\,,
\end{align}
together with
\begin{align}
\pl^\pm x^\mp&=g^{\pm \mp}\,,& \pl\, \bar{z}&=g^{z\bar{z}}\,,& \bar{\pl}\,
z&=g^{\bar{z}z}\,,
\end{align}
where we fix the line-element as
\begin{equation}
ds^2=-2 dx^+ dx^-+2dz d\bar{z}\,.
\end{equation}
Using an analogous decomposition for the auxiliary variables $u^\mu=(u^+,u^-,u,\bar{u})$ one can first rewrite the Fierz system as:
\begin{subequations}
\begin{align}
(-2\pl^+\pl^-+2\pl\bar{\pl})\Phi(x,u)&=0\,,\\
(-\pl^+\pl_u^--\pl^-\pl_u^++\pl\bar{\pl}_u+\bar{\pl}\pl_u)\Phi(x,u)&=0\,,\\
(-2\pl_u^+\pl_u^-+2\pl_u\bar{\pl}_u)\Phi(x,u)&=0\,,
\end{align}
\end{subequations}
and fix completely the leftover gauge symmetries:
\begin{equation}
\delta_\epsilon \Phi(x,u)=(-u^+\pl^--u^-\pl^++u\bar{\pl}+\bar{u}\pl)\,\epsilon\,,
\end{equation}
by requiring $\pl_u^+\,\Phi(x,u)=0$. One thus has
\begin{subequations}
\begin{align}
(-2\pl^+\pl^-+2\pl\bar{\pl})\Phi(x,u)&=0\,,\\
\pl_u^-\Phi(x,u)&=\frac1{\pl^+}(\pl\,\bar{\pl}_u+\bar{\pl}\,\pl_u)\Phi(x,u)\,,\\
\pl_u\bar{\pl}_u\,\Phi(x,u)&=0\,,
\end{align}
\end{subequations}
ending up with the two physical helicities in 4d:
\begin{equation}
    \Phi(x,u)=\sum_s{\varphi}_{-s}(x)\,u^s+\varphi_{+s}(x)\,\bar{u}^s\equiv\bar{\varphi}(x,u)+\varphi(x,\bar{u})\,.
\end{equation}
These form a pair of complex conjugate scalar fields ${\varphi}_{-s}(x)\equiv \Phi_{\bar{z}(s)}(x)$ and $\varphi_s(x)\equiv \Phi_{{z}(s)}(x)$.

The light-cone gauge for the couplings (at any order) can be achieved by the following replacements
\begin{subequations}\label{light-cone}
\begin{align}
Y_{ij}&\rightarrow -(\pl_{x_i}^+)^{-1}\left[\bar{P}_{ij}\pl_{u_1}+P_{ij}\bar{\pl}_{u_1}\right]\,,\\
H_{ij}&\rightarrow +\pl_{u_i}\bar{\pl_{u_j}}+\pl_{u_j}\bar{\pl_{u_i}}\,,
\end{align}
\end{subequations}
with the notation $P_{ij}=\pl_i\,\pl^+_j-\pl_j\,\pl^+_i$.
At cubic order it is also useful to recall the following reduction of the cubic structure $G$ of appendix \ref{cubic}:
\begin{subequations}
\begin{align}
G&\rightarrow\bar{\pl}_{u_1}\bar{\pl}_{u_2}{\pl}_{u_3}\,\left(\frac{\pl_{x_3}^+}{\pl_{x_1}^+\pl_{x_2}^+}\,P\right)+{\pl}_{u_1}{\pl}_{u_2}\bar{\pl}_{u_3}\,\left(\frac{\pl_{x_3}^+}{\pl_{x_1}^+\pl_{x_2}^+}\,\bar{P}\right)+\text{cyclic}\,,
\end{align}
\end{subequations}
where we have introduced the following two (anti-)holomorphic light-cone momenta which are useful in the $n=3$ case:
\begin{subequations}
\begin{align}
P&=\frac13\left[\pl_{x_1}\,(\pl^+_{x_2}-\pl^+_{x_3})+\pl_{x_2}\,(\pl^+_{x_3}-\pl^+_{x_1})+\pl_{x_3}\,(\pl^+_{x_1}-\pl^+_{x_2})\right]\,,\\
\bar{P}&=\frac13\left[\bar{\pl}_{x_1}\,(\pl^+_{x_2}-\pl^+_{x_3})+\bar{\pl}_{x_2}\,(\pl^+_{x_3}-\pl^+_{x_1})+\bar{\pl}_{x_3}\,(\pl^+_{x_1}-\pl^+_{x_2})\right]\,.
\end{align}
\end{subequations}
Using the above notation one can write the most general light-cone cubic solution as
\begin{multline}
    S^{(3)}=\sum_{s_1,s_2,s_3}\bar{g}_{s_1,s_2,s_3}\,P^{-s_1-s_2-s_3}(\pl^+_1)^{s_1}(\pl^+_1)^{s_2}(\pl^+_1)^{s_3}\\+g_{s_1,s_2,s_3}\bar{P}^{s_1+s_2+s_3}(\pl^+_1)^{-s_1}(\pl^+_1)^{-s_2}(\pl^+_1)^{-s_3}\,,
\end{multline}
where the sums range over helicities $s_i=\pm|s_i|$ and we have hidden the dependence on $\pl_u^{s}$ and $\bar{\pl}_u^s$ as it is redundant with the helicity.

Locality in the light-cone is simply the statement that no inverse $P$ is present (the coupling constant should vanish accordingly), while the main difference with respect to the covariant classification is the presence of one structure for each choice of helicities: $+++$, $+--$, $++-$ and $---$ (see \cite{Conde:2016izb,Sleight:2016xqq}).

The light-cone formalism turns out to be particularly useful to compute current exchanges, since the form of the propagator drastically simplifies as a consequence of the identities:
\begin{align}
    u^2&=2\bar{u}u\,,&v^2&=2\bar{v}v\,,& u\cdot v=u\bar{v}+\bar{u}v\,.
\end{align}
Plugging the above light-cone reduction expressions into the formula \eqref{4dprop} we then arrive at the very simple expression:
\begin{align}
\mathcal{P}_r(u,v)&=-\frac{1}{p^2}\,\frac1{r!^2}\left(u^r\bar{v}^r+\bar{u}^r{v}^r\right)\,,
\end{align}
for the propagator. In general the current exchange in a given channel is then made of 4 pieces and reads:
\begin{align}\label{lcprop}
\mathcal{E}_{\sf s}=\frac{1}{{\sf s}}\,\sum_{r}\Big[\bar{g}_{s_1,s_2,r}&\,\bar{g}_{-r,s_3,s_4}P_{12}^{-s_1-s_2-r}P_{34}^{-s_3-s_4+r}(\pl_1^+)^{s_1}(\pl_2^+)^{s_2}(\pl_3^+)^{s_3}(\pl_4^+)^{s_4}\\
+{g}_{s_1,s_2,r}&\,{g}_{-r,s_3,s_4}\bar{P}_{12}^{s_1+s_2+r}\bar{P}_{34}^{s_3+s_4-r}(\pl_1^+)^{-s_1}(\pl_2^+)^{-s_2}(\pl_3^+)^{-s_3}(\pl_4^+)^{-s_4}\nonumber\\
+\bar{g}_{s_1,s_2,r}\,{g}_{-r,s_3,s_4}&P_{12}^{-s_1-s_2-r}\bar{P}_{34}^{s_3+s_4-r}(\pl_1^+)^{s_1}(\pl_2^+)^{s_2}(\pl_1^++\pl_2^+)^{2 r}(\pl_3^+)^{-s_3}(\pl_4^+)^{-s_4}\nonumber\\
+{g}_{s_1,s_2,r}\,\bar{g}_{-r,s_3,s_4}&\bar{P}_{12}^{s_1+s_2+r}{P}_{34}^{-s_3-s_4+r}(\pl_1^+)^{-s_1}(\pl_2^+)^{-s_2}(\pl_1^++\pl_2^+)^{-2r}(\pl_3^+)^{s_3}(\pl_4^+)^{s_4}\Big]\,,\nonumber
\end{align}
where $s_i$ and $r$ are helicities that can take values $s_i=\pm|s_i|$ and $r=\pm|r|$ while the power of $P_{ij}$ and $\bar{P}_{ij}$ must be positive and corresponding sum over $r=\pm|r|$ is constrained accordingly. As shown by Metsaev in \cite{Metsaev:1991nb}, and as can be easily verified by explicit computation, fixing the coefficients to be
\begin{align}\label{metcho}
g_{s_1,s_2,s_3}&=\frac1{\Gamma(s_1+s_2+s_3)}\,,& \bar{g}_{s_1,s_2,s_3}&=\frac1{\Gamma(-s_1-s_2-s_3)}\,,
\end{align}
and summing over the poles contributing to a planar structure $\mathcal{E}_{1234}$ (the permutation $\{1,2,3,4\}$ and $\{4,1,2,3\}$), gives a vanishing contribution from the terms of the type $\bar{g}\bar{g}$ and $gg$. This property defines Metsaev's solution both with and without Chan-Paton factors since the cancellation arises at the level of the single planar exchange $\mathcal{E}_{1234}$. It is interesting to stress that the latter cancellation can only be achieved due to the lower derivative exotic structures present in the light-cone formalism, which do not admit a standard covariant form \cite{Conde:2016izb,Sleight:2016xqq}. On the other hand, the actual value of the couplings is in accordance with a generalised version of minimal coupling to gravity and to HS, as shown in \cite{Sleight:2016xqq,Ponomarev:2016lrm}. 

An explicit verification of the above statement follows from the fact that, upon fixing the coupling to those of Metsaev \eqref{metcho}, the above $gg$ and $\bar{g}\bar{g}$ infinite series \eqref{lcprop} truncates to finite sums at fixed external spins. These can be re-summed explicitly as:
\begin{align}
O(gg)&\sim \frac1{{\sf s}}\,(\pl_1^+)^{-s_1}(\pl_2^+)^{-s_2}(\pl_3^+)^{-s_3}(\pl_4^+)^{-s_4}\,\bar{P}_{12}\bar{P}_{34}(\bar{P}_{12}+\bar{P}_{34})^{s_1+s_2+s_3+s_4-2}\,,
\end{align}
with analogous results for $\bar{g}\bar{g}$ terms. Using momentum conservation, the above easily allows to compute the full planar particular solution to be given by $g\bar{g}$ contributions only:
\begin{align}\label{exchstu}
\mathcal{E}_{1234}=\frac{1}{{\sf s}}\,\sum_{r}\Big[(\pl_1^+)^{s_1}(\pl_2^+)^{s_2}&(\pl_1^++\pl_2^+)^{2 r}(\pl_3^+)^{-s_3}(\pl_4^+)^{-s_4}\\
&\times\frac{1}{\Gamma(-s_1-s_2-r)\Gamma(s_3+s_4-r)}P_{12}^{-s_1-s_2-r}\bar{P}_{34}^{s_3+s_4-r}\Big]\nonumber\\
+\frac{1}{{\sf u}}\,\sum_{r}\Big[(\pl_1^+)^{s_1}(\pl_2^+)^{s_2}&(\pl_4^++\pl_1^+)^{2 r}(\pl_3^+)^{-s_3}(\pl_4^+)^{-s_4}\nonumber\\
&\times\frac{1}{\Gamma(-s_4-s_1-r)\Gamma(s_2+s_3-r)}P_{41}^{-s_4-s_1-r}\bar{P}_{23}^{s_2+s_3-r}\Big]\nonumber\\
+\frac{1}{{\sf s}}\,\sum_{r}\Big[(\pl_1^+)^{-s_1}(\pl_2^+)^{-s_2}&(\pl_1^++\pl_2^+)^{-2 r}(\pl_3^+)^{s_3}(\pl_4^+)^{s_4}\nonumber\\
&\times\frac{1}{\Gamma(s_1+s_2+r)\Gamma(-s_3-s_4+r)}\bar{P}_{12}^{s_1+s_2+r}{P}_{34}^{-s_3-s_4+r}\Big]\nonumber\\
+\frac{1}{{\sf u}}\,\sum_{r}\Big[(\pl_1^+)^{-s_1}(\pl_2^+)^{-s_2}&(\pl_4^++\pl_1^+)^{-2 r}(\pl_3^+)^{s_3}(\pl_4^+)^{s_4}\nonumber\\
&\times\frac{1}{\Gamma(s_4+s_1+r)\Gamma(-s_2-s_3+r)}\bar{P}_{41}^{s_4+s_1+r}{P}_{23}^{-s_2-s_3+r}\Big]\nonumber
\end{align}
Let us stress that for fixed external spins the above sums do not truncate to a finite number of exchanges and the series is truly infinite as opposed to the $gg$ and $\bar{g}\bar{g}$ parts of the exchange. On the other hand, requiring the cancellation of all $gg$ and $\bar{g}\bar{g}$ poles is sufficient to fix all cubic coupling constants. In a sense, the main simplification of the light-cone formalism is precisely the possibility of disentangling these two contributions: $g\bar{g}$ vs. $gg$, $\bar{g}\bar{g}$, studying them separately and setting the latter to zero (as they would contribute a non-local term to the solution for the contact term). In covariant language these contribution are mixed, and it is not possible to solve the cancellation of $gg$ poles separately from $g \bar{g}$ poles (see however appendix \ref{s1s1} for analysis of more general couplings in covariant formalism requiring the maximal cancellations of the non-localities).

Focusing on the $s_1$-$0$-$0$-$0$ case it is interesting to compare more in detail the form of the exchange above with the covariant formalism discussed previously in section \ref{lookingfor}. Choosing, for definiteness, a positive helicity $+|s_1|$ we get lower derivative exchanges only from the last two terms in \eqref{exchstu}, as the first two terms are non-vanishing only if $r<-|s_1|$:
\begin{align}
\mathcal{E}_{1234}^{(+|s_1|,0,0,0)}=&\frac{1}{{\sf s}}\,\sum_{r}\Big[(\pl_1^+)^{s_1}(\pl_1^++\pl_2^+)^{2 r}\frac{1}{\Gamma(-s_1-r)\Gamma(-r)}P_{12}^{-s_1-r}\bar{P}_{34}^{-r}\Big]\nonumber\\
+&\frac{1}{{\sf u}}\,\sum_{r}\Big[(\pl_1^+)^{s_1}(\pl_4^++\pl_1^+)^{2 r}\frac{1}{\Gamma(-s_1-r)\Gamma(-r)}P_{41}^{-s_1-r}\bar{P}_{23}^{-r}\Big]\nonumber\\
+&\frac{1}{{\sf s}}\,\sum_{r}\Big[(\pl_1^+)^{-s_1}(\pl_1^++\pl_2^+)^{-2 r}\frac{1}{\Gamma(s_1+r)\Gamma(r)}\bar{P}_{12}^{s_1+r}{P}_{34}^{r}\Big]\nonumber\\
+&\frac{1}{{\sf u}}\,\sum_{r}\Big[(\pl_1^+)^{-s_1}(\pl_4^++\pl_1^+)^{-2 r}\frac{1}{\Gamma(s_1+r)\Gamma(r)}\bar{P}_{41}^{s_1+r}{P}_{23}^{r}\Big]\,.
\end{align}
If we restrict to the bad contributions with $|r|<|s_1|-1$ (i.e. those poles which could not be cancelled in the covariant formalism, see section \ref{lookingfor} equation \eqref{dangerous}), we then arrive to:\footnote{One can also go a little further for the sum over $r$, resumming the above series in terms of Bessel functions using the identity:
\begin{align}
    \sum_r\frac{x^{\alpha+r}y^{-\beta+r}}{\Gamma(\alpha+r)\Gamma(-\beta+r)}&=x^{\alpha+\beta}(xy)^{-\tfrac{\alpha+\beta}2+1}I_{\alpha+\beta}(2\sqrt{xy})\,,& \alpha>0\,.
\end{align}}
\begin{align}\label{exchlcbis}
\mathcal{E}_{1234}^{(+|s_1|,0,0,0)}=\frac{1}{{\sf s}}\,\sum_{r=1}^{s_1-2}\Big[(\pl_1^+)^{-s_1}&(\pl_1^++\pl_2^+)^{-2 r}\frac{1}{\Gamma(s_1+r)\Gamma(r)}\bar{P}_{12}^{s_1+r}{P}_{34}^{r}\Big]\nonumber\\
+\frac{1}{{\sf u}}\,\sum_{r=1}^{s_1-2}\Big[(\pl_1^+)^{-s_1}&(\pl_4^++\pl_1^+)^{-2 r}\frac{1}{\Gamma(s_1+r)\Gamma(r)}\bar{P}_{41}^{s_1+r}{P}_{23}^{r}\Big]+\ldots\,.
\end{align}
Observe that all contributions from the exotic coupling $s_1$-$r$-$0$ with $s_1-r$ derivatives cancel out in the $gg$ part of the exchange, and are indeed a key ingredient for that cancellation. However, there is still a leftover dangerous contribution which is the \emph{same} contribution one recovers in covariant form from the cubic couplings $s_1$-$r$-$0$ with $s_1+r$ derivatives which we have discussed in eq.~\eqref{dangerous}.

To conclude, we stress that all the dangerous poles present in the covariant formulation (see section \ref{lookingfor}) are \emph{also} present in the above exchange amplitudes, and cannot be cancelled by an appropriate homogeneous solution. A way out could be provided by a homogeneous solution which does not exist in the covariant classificaton (see section \ref{Homosol}), in analogy to the extra cubic structures which appear on the light-cone \cite{Bengtsson:1983pd,Bengtsson:1986kh} compared to the covariant classification \cite{Metsaev:2005ar,Manvelyan:2010jr,Sagnotti:2010at}. However, as we discuss below, the existence of such a structure does not seem feasible.

To reiterate, from the perspective of the Noether procedure, Metsaev's solution can be considered as the solution obtained by requiring the maximal cancellation of poles in the exchange (which was referred to as minimal scheme in \cite{Taronna:2011kt}). As anticipated, however this solution still leaves some leftover pole contributions that would be required to be cancelled by a judicious choice of the homogeneous solution.\footnote{The above results also imply that fixing either $g=0$ or $\bar{g}=0$ for all spins, Metsaev's solution \cite{Metsaev:1991nb,Metsaev:1991mt} provides a full quartic solution to Noether procedure, as mentioned in \cite{Ponomarev:2016lrm}. This follows because all non-local contributions to the particular solution are set to zero identically and no homogeneous solution is then required to cancel any pole factor. This type of solution parallels the solution discussed in \eqref{trivialS} and is local simply because the coefficients of the non-local obstructions vanish identically.} To investigate this issue more closely, one may consider, for instance, the light-cone reduction of the covariant and Poincar\'e invariant homogeneous solution we have discussed in the $s$-$0$-$0$-$0$ case. It is useful to recall that the Mandelstam variables on the light-cone read:
\begin{equation}
{\sf s}=-\frac{2}{\pl_1^+\,\pl_2^+}\,P_{12}\bar{P}_{12}\,.
\end{equation}
In the $s$-$0$-$0$-$0$ case only one covariant building block is present \eqref{NoetherSol} (up to Mandelstam invariants), which gives rise to the following light-cone Poincar\'e invariant quartic structures:
\begin{align}\label{s000lc}
{\sf u}\,\pl_{u_1}\cdot p_2-{\sf s}\,\pl_{u_1}\cdot p_4&\ \longrightarrow\  {\sf s\,u}\,\left[\pl_{u_1}\left(\frac{\pl_2^+}{P_{12}}+\frac{\pl_4^+}{P_{14}}\right)+\bar{\pl}_{u_1}\left(\frac{\pl_2^+}{\bar{P}_{12}}+\frac{\pl_4^+}{\bar{P}_{14}}\right)\right]\,,
\end{align}
whose powers generate the HS light-cone structures.
The above gives one structure of the type $(+s)$-$0$-$0$-$0$ and one structure of the type $(-s)$-$0$-$0$-$0$, which however are not enough to cancel all pole factors present in \eqref{exchlcbis}, as discussed in section \ref{lookingfor}.

Let us stress that at the cubic order in both light-cone and covariant formalism there is a unique structure for the $\pm s$-$0$-$0$ case which is nothing but the $s$-th power of the $\pm1$-$0$-$0$ structure. Indeed, the light-cone gauge gives precisely one structure for each helicity combination (up to possible Mandelstam invariants). If this continues to be true at higher orders (consistently with the analysis of \cite{Bengtsson:2016hss}) this means that also at 4pt level there will be one structure $\pm s$-$0$-$0$-$0$ for the $s$-$0$-$0$-$0$ case, which must then coincide with \eqref{s000lc}.
To conclude, we are led to argue that there is no additional structure in the light-cone on top of the above covariant ones. Therefore, the non-local obstruction appears to persists also working on the light cone.

\section{Summary and conclusions}

To summarise, in this note we aimed to provide some details and clarification on the obstructions to interacting HS theories in flat space:

We have expanded on the results of \cite{Taronna:2011kt}, for the simplest example quartic vertices in flat space with a single HS external leg and three scalars. We have shown that there are $1/\Box$ obstructions to locality proportional to bi-linears in the cubic coupling coefficients of the type $g_{s_1,r,0}\,g_{r,0,0}$ with $r\leq s_1-2$ or $r\leq s_1-4$ for the color non-singlet and singlet sector respectively. These obstructions are labelled by the exchanged spin $r$ and are a finite number for given external spin $s_1$. In particular no obstruction arises for $s_1<2$. Such obstructions appear to be generic features of quartic couplings with higher-spin external legs and are present also for more general quartic interactions (see Appendix \ref{s1s1} and \cite{Taronna:2011kt}). 

Owing to the fact that the above obstructions have been uncovered using a manifestly covariant language, we have also discussed the full light-cone exchange amplitude in 4d. We have shown that it is possible to recover Metsaev's solution (chiral and non) from the requirement of maximal cancellation of non-localities (referred as minimal scheme in \cite{Taronna:2011kt}).
We have also argued that the non-local obstruction originally found in \cite{Taronna:2011kt} continues to be present also in the light-cone formalism.

\section*{Acknowledgments}
\label{sec:Aknowledgements}

I am indebted to Charlotte Sleight for very useful discussions, comments and carefully reading this manuscript. I would also like to thank G. Barnich and especially R. Roiban and A. Tseytlin for discussions, comments and for sharing their draft \cite{Roiban:2017iqg} prior to submission. The research of M. Taronna is partially supported by the Fund for Scientific Research-FNRS Belgium, grant FC 6369 and by the Russian Science Foundation grant 14-42-00047 in association with Lebedev Physical Institute.

\begin{appendix}
\section{Notation and conventions}\label{not}
In this paper we use generating function notation for the fields:
\begin{equation}
\phi(x,u)=\frac1{s!}\phi_{\mu(s)}(x)u^{\mu(s)}\,,
\end{equation}
in terms of an auxiliary variable $u^\mu$. Within this formalism it is also convenient to introduce scalar contractions between the auxiliary variables and ingoing momenta as
\begin{align}
Y_{ij}&=-i\pl_{u_i}\cdot p_j=\pl_{u_i}\cdot\pl_{x_j}\,,& H_{ij}&=\pl_{u_i}\cdot \pl_{u_j}\,,& {\sf s}_{ij}&=-(p_i+p_j)^2=(\pl_{x_i}+\pl_{x_j})^2\,.
\end{align}
A generic multilinear functional of the fields can then be encoded into a function of the above building blocks of the type
\begin{equation}
\mathcal{A}^{(n)}[\phi_i]=f^{(n)}(Y_{ij},H_{ij},{\sf s}_{ij})\,\phi_{1}(x_1,u_1)\cdots\phi_n(x_n,u_n)\Big|_{x_i=x,u_i=0}\,.
\end{equation}

\section{The cubic coupling covariant classification}\label{cubic}

The flat space cubic coupling classification in covariant form can be easily summarised in using generating functions. The transverse and traceless part of the cubic couplings in flat space can be written in covariant form as general functions of 4 building blocks which we can arrange as:
\begin{equation}\label{cub}
f^{[k]}_{s_1,s_2,s_3}=Y_1^{s_1-k}Y_2^{s_2-k}Y_3^{s_3-k}G^k\,,
\end{equation}
where we drop the suffix $[k]$ when no ambiguity can arise (as in the $s_1$-$s_2$-$0$ case where $k\equiv0$).
Above we have used
\begin{subequations}
\begin{align}
Y_1&=\pl_{u_1}\cdot p_2\,,& Y_2&=\pl_{u_2}\cdot p_3\,,& Y_3&=\pl_{u_3}\cdot p_1\,,\\
H_1&=\pl_{u_2}\cdot \pl_{u_3}\,,& H_2&=\pl_{u_3}\cdot \pl_{u_1}\,,& H_3&=\pl_{u_1}\cdot \pl_{u_2}\,,
\end{align}
\end{subequations}
and
\begin{equation}
G=Y_1 H_1+Y_2H_2+Y_3H_3\,.
\end{equation}
The most general covariant and local cubic coupling can then be written as a generic sum over the spins:
\begin{equation}
S^{(3)}=\int\sum_{s_1,s_2,s_3,k} g_{s_1,s_2,s_3}^{[k]}\,f^{[k]}_{s_1,s_2,s_3}\,,
\end{equation}
where we have introduced arbitrary coupling constants $g_{s_1,s_2,s_3}^{[k]}$.
In 4d only two couplings survive, for $k=0$ and $k=s_{\text{min}}$ due to the Schouten identity:
\begin{equation}
Y_1\,Y_2\,Y_3\, G\sim p_i^2\,.
\end{equation}
The generating function formalism also allows to easily extend the above TT discussion to the de Donder gauge, which is necessary for higher-point computations. At quartic order it is enough to consider the de-Donder completion of a vertex which has two on-shell legs (say $\phi_2$ and $\phi_3$). 

In the de Donder gauge the equations of motion for the fields read:
\begin{subequations}
\begin{align}
\Box\phi&=0\,,&\Box\phi^\prime&=0\,,& \pl_u\cdot\pl_x\phi&=\frac12\,u\cdot\pl_x\phi^\prime\,,\\
\delta^{(0)}_\epsilon\phi&=u\cdot\pl_x\epsilon\,,& \delta^{(0)}_\epsilon\phi^\prime&=2\,\pl_u\cdot\pl_x\epsilon\,.
\end{align}
\end{subequations}
One can then show that the TT vertex with 2 on-shell external legs and one leg in the de-Donder gauge is exact and can be used for 4pt computations without any loss of generality. The gauge variation of the vertex with respect to the on-shell leg $\phi_1$ with $\phi_3$ in the de Donder gauge reads:
\begin{equation}\label{def1}
\delta^{(0)}_\epsilon f(Y_i,G)=\frac12(\pl_{Y_1}f(Y_i,G)\,\epsilon_1\,\phi_2)\Box_3\phi_3\equiv T_{12}\epsilon_1\phi_2\,\Box_3\phi_3\,.
\end{equation}
The gauge variation of with respect to the on-shell leg $\phi_1$ with now $\phi_2$ in the de Donder gauge instead reads:
\begin{equation}\label{def2}
\delta_\epsilon^{(0)}f(Y_i,G)=\frac12\left[(\pl_{Y_1}+\pl_{u_2}^2\pl_{H_3}\pl_{Y_2})f(Y_i,G)\,\epsilon_1\,\phi_3\right]\Box_2\phi_2\equiv T_{13}\epsilon_1\phi_3\,\Box_2\phi_2\,.
\end{equation}
From the above, one can rewrite the variation of the cubic vertex in terms of the deformation of the gauge transformations taking into account that the de Donder form of the Fronsdal kinetic term reads:
\begin{equation}
S^{(2)}=\frac{s!}2\int\phi(\pl_u)\left(1-\frac14 u^2\pl_u^2\right)\Box\,\phi(u)\Big|_{u=0}\,.
\end{equation}
Therefore we arrive at
\begin{equation}\label{gaugedef}
\delta^{(0)}_\epsilon f=-\,\delta_{\epsilon}^{(1)}\phi(\pl_u)(1-\frac14 u^2\pl_u^2)\Box\,\phi(u)\,.
\end{equation}

For the computation of the exchange we need to extract from the above couplings appropriately improved conserved currents. To this end we introduce the following building blocks:
\begin{align}
\bar{Y}_1&=\pl_{u_1}\cdot\pl_{x_2}\,,&\bar{Y}_2&=-\pl_{u_2}\cdot\pl_{x_1}\,,& \bar{Y}_{3}&=\frac12\pl_{u_3}\cdot\pl_{x_{12}}\,,
\end{align}
obtained from the above ones up to integrations by parts.
One can then construct on-shell conserved currents simply replacing $Y_i$ with $\bar{Y}_i$, which can then be employed to compute exchange amplitudes.

Before concluding this section it is useful to compute the $n$-th trace of the currents with respect to the leg attached to the propagator, while the other two legs are kept on-shell. This can be easily obtained using the formula:
\begin{equation}
[f(\bar{Y}_i,G)\,,u_3^2]_{u_3=0}=-\left[Y_1\,Y_2\,\pl_{Y_3}\,\pl_{G}+\frac{{\sf s}}{4}\left(H_3^2\pl_G^2+2H_3\pl_{Y_3}\pl_G+\pl_{Y_3}^2\right)\right]\,f(Y_i,G)\,.
\end{equation}
The above implies that if a coupling is not proportional to $G$ then its trace is proportional to ${\sf s}$ and so it does not contribute to the pole part of the exchange amplitude. We can apply the above formula to $f^{[k]}_{s_1,s_2,s_3}$ obtaining:
\begin{align}\label{traceC}
[f^{[k]}_{s_1,s_2,s_3},(u_3^2)^n]&=(-1)^n\,\frac{\Gamma(s_3-k+1)}{\Gamma(s_3-k-n+1)}\frac{\Gamma(k+1)}{\Gamma(k-n+1)}\,f_{s_1,s_2,s_3-2n}^{[k-n]}\nonumber\\&\equiv(-1)^n\,A_{s_3,k}^{[n]}\,f_{s_1,s_2,s_3-2n}^{[k-n]}+O({\sf s})\,.
\end{align}
In 4d there are further simplifications because $f^{[k]}_{s_1,s_2,s_3}$ is not proportional to $p_i^2$ only if $k=0$ and $k=s_{\text{min}}$. This means that the latter traces will not contribute to the non-local part of the exchange being either vanishing on-shell or proportional to ${\sf s}$. In this case there are 2 relevant cases:
\begin{itemize}
\item $s_3=s_{\text{min}}$: In this case $k=s_3$ and then $s_3-n>s_3-2n$ so that the coupling is effectively traceless.
\item $s_3\neq s_{\text{min}}$: In this case let us fix $s_1=s_{\text{min}}$. For any $n$ such that $s_3-2n\geq s_1$ each trace will not contribute to the non-local part of the exchange in 4d. If instead $s_3-2n < s_1$ then the corresponding trace would contribute in 4d only if $s_1-n=s_3-2n$ which fixes $n=\text{min}(s_3-s_{\text{min}},s_{\text{min}})$. We thus find at most one trace contributing to the non-local part.
\end{itemize}





\section{Yang-Mills Noether procedure}\label{YMnoether}

In this appendix we employ our method to find YM couplings enforcing locality. The starting point is the generating function of two derivative cubic couplings among scalars and gauge bosons:
\begin{equation}
\mathcal{V}_{123}=g_{111}f_{111}^{[1]}+g_{001}f_{001}^{[0]}+g_{010}f_{010}^{[0]}++g_{100}f_{100}^{[0]}\,.
\end{equation}
The particular solution to the Noether procedure is then given by minus the current exchange amplitude whose color ordered representative reads:
\begin{equation}
\mathcal{E}_{1234}=-\frac1{{\sf s}}\mathcal{V}_{12u}\mathcal{V}_{v34}\, (1+u\cdot v)-\frac1{{\sf u}}\mathcal{V}_{41u}\mathcal{V}_{v23}\, (1+u\cdot v)\,.
\end{equation}
where the $(1+u\cdot v)$ gives the sum of scalar and gauge boson propagator.
The gauge transformations of the exchange can then be computed using:
\begin{align}
\delta^{(0)}_4\mathcal{V}_{v34}&=+\tfrac12 g_{111}\left[ Y_{34}\,\pl_{v}\cdot(\pl_{x_3}+\pl_{x_4})-\pl_{v}\cdot\pl_{u_3}\,{\sf s}\right]-\tfrac12 g_{001}\,{\sf s}\,,\\
\delta^{(0)}_4\mathcal{V}_{41u}&=-\tfrac12 g_{111}\left[ Y_{14}\,\pl_{u}\cdot(\pl_{x_4}+\pl_{x_1})-\pl_{u}\cdot\pl_{u_1}\,{\sf u}\right]+\tfrac12 g_{100}\,{\sf u}\,.
\end{align}
All in all one ends up with the following local gauge variation:
\begin{align}
\delta_4^{(0)}\mathcal{E}_{1234}=&-\frac14\,g_{111}^2\left(H_{12}Y_{34}+H_{23}Y_{14}-2H_{31}Y_{24}\right)\\&\nonumber-\frac14 g_{100}\left(g_{111}\pl_{u_1}\cdot\pl_{x_{23}}-2g_{001}\,Y_{12}\right)
+\frac14 g_{001}\left(g_{111}\pl_{u_3}\cdot\pl_{x_{12}}+2g_{100}\,Y_{32}\right)\\&\nonumber-\frac12 g_{010}\left(g_{100}\,Y_{23}+g_{001}\,Y_{21}\right)\,.
\end{align}
Combining the above with gauge variations with respect to the other external legs and requiring that the above can be integrated to a a local quartic coupling (namely that up to integrations by parts the above is proportional to $\pl_{x_4}$) forces $g_{111}=g_{001}=g_{010}=g_{100}=g$ so that one finds all quartic contact terms in YM Lagrangian together with the condition relating cubic and quartic coupling coefficients (charge conservation):
\begin{multline}
\mathcal{V}_{1234}=-\frac14\,g^2\left(H_{12}H_{34}+H_{23}H_{14}-2H_{31}H_{24}\right)\\-\frac14 g^2\left(H_{14}+H_{23}\right)
-\frac14 g^2\left(H_{34}+H_{12}\right)+\frac12 g^2\left(H_{24}+H_{13}\right)
\end{multline}
The latter quartic coupling precisely accounts for the $\phi^2 A^2$ and $A^4$ contact terms in YM theory once all color orderings are combined together.

\section{Current exchanges}

\subsection{Four dimensions}\label{currexch}

The propagator polynomial simplifies in 4d and can be expressed in terms of Chebyshev polynomials as:
\begin{equation}\label{4dprop}
\mathcal{P}( u,v)=-\frac{i}{p^2}\,\frac{2}{(r!)^2}\,\left(\frac{u^2\,v^2}{4}\right)^{r/2}T_r\left(\frac{u\cdot v}{\sqrt{u^2\,v^2}}\right)\,.
\end{equation}
Due to the simplifications that arise in 4d, in this section we consider the following generic ${\sf s}$-channel exchange:
\begin{multline}
\mathcal{E}(s_1,s_2|r|s_3,s_4)=g_{s_1,s_2,r}^{[k]}g_{r,s_3,s_4}^{[q]}\frac{1}{{\sf s}}\,\Big[f_{s_1,s_2,r}^{[k]}\,f_{r,s_3,s_4}^{[q]}\,\frac{\left(u\cdot v\right)^r}{(r!)^2}\\+\sum_{n>0}2^{-r+1}t_{r,n}\,A_{r,k}^{[n]}\,A_{r,q}^{[n]}\,f_{s_1,s_2,r-2n}^{[k-n]}f_{r-2n,s_3,s_4}^{[q-n]}\frac{\left(u\cdot v\right)^{r-2n}}{(r-2n)!^2}\Big]+O({\sf s})\,,
\end{multline}
where $t_{r,n}$ the coefficients in the expansion of the Chebyshev polynomial multiplied by $\frac{(r-2n)!^2}{r!^2}$. Notice that the trace parts contribute to the pole only if both side of the exchange are tracefull on-shell. If $r>s_i$ this happens only if $q=k$ which means that $\text{min}(s_1,s_2)=\text{min}(s_3,s_4)$.

In the following we will complete the computation of the exchange simply evaluating $f_{s_1,s_2,r}^{[k]}\,f_{r,s_3,s_4}^{[q]}\tfrac{\left(u\cdot v\right)^r}{r!^2}$. The corresponding result can be easily extracted and reads:
\begin{multline}
f_{s_1,s_2,r}^{[k]}\,f_{r,s_3,s_4}^{[q]}\frac{\left(u\cdot v\right)^r}{r!^2}=(\pl_{u_1}\cdot p_2)^{s_1-k}(-\pl_{u_2}\cdot p_1)^{s_2-k}(\pl_{u_3}\cdot p_4)^{s_3-q}(-\pl_{u_4}\cdot p_3)^{s_4-q}\\\sum_{\ga=0}^q
\binom{k}{\ga}\binom{r}{q}^{-1}
\binom{r-k}{q-\alpha}
\,(G_{12r}\cdot G_{r34})^\ga(\tfrac12 p_{12}\cdot G_{r34})^{q-\ga}(\tfrac12 G_{12r}\cdot p_{34})^{k-\ga}(\tfrac{t-u}4)^{r-q-k+\ga}\,,
\end{multline}
where for convenience we have defined
\begin{equation}
G_{12r}^\mu\equiv \pl_{\pl_{u_r}}^\mu G_{12r}=\pl_{u_1}\cdot p_2\, \pl_{u_2}^\mu-\pl_{u_2}\cdot p_1\, \pl_{u_1}^\mu+\frac12 p_{12}^\mu\,\pl_{u_1}\cdot\pl_{u_2}\,.
\end{equation}

Various simplifications arise in the above exchange when considering particular process in 4d (in the following we restrict for ease of notation to the non-local part in the exchange:
\begin{itemize}
\item $s_1000$: Since all couplings do not depend on $G$ all trace components drop out (this is true in any dimension).
\begin{equation}\label{Exchs000}
\mathcal{E}_{\sf s}=-\frac{1}{{\sf s}}\,g_{s_10r} \,g_{r00}\,(\pl_{u_1}\cdot p_2)^{s_1}\left(\tfrac{{\sf t}-{\sf u}}4\right)^r\,.
\end{equation}
\item $s_1s_200$: Since one never finds two couplings both depending on $G$ again all trace components of the coupling drop out.
\begin{equation}\label{Exchs1s200}
\mathcal{E}_{\sf s}=-\frac{1}{{\sf s}}\,g^{[k]}_{s_1s_2r} \,g_{r00}\,(\pl_{u_1}\cdot p_2)^{s_1-k}(-\pl_{u_2}\cdot p_1)^{s_2-k}\\(G_{12r}\cdot p_3)^{k}\left(\tfrac{{\sf t}-{\sf u}}{4}\right)^{r-k}\,.
\end{equation}
\item $s_1s_2s_30$: Since again one never finds two couplings both depending on $G$ again all trace components of the coupling drop out.
\begin{equation}
\mathcal{E}_{\sf s}=-\frac{1}{{\sf s}}\,g^{[k]}_{s_1s_2r} \,g_{rs_30}\,(\pl_{u_1}\cdot p_2)^{s_1-k}(-\pl_{u_2}\cdot p_1)^{s_2-k}(\pl_{u_3}\cdot p_4)^{s_3}\\(G_{12r}\cdot p_3)^{k}\left(\tfrac{{\sf t}-{\sf u}}{4}\right)^{r-k}\,.
\end{equation}
\end{itemize}
The above result allows in principle to extend the $s_1$-$0$-$0$-$0$ analysis to more general cases as we show in the next section. Since anyway the non-localities we find in the $s_1$-$0$-$0$-$0$ case cannot be compensated in covariant language we do not repeat the analysis of the more general $s_1$-$s_2$-$s_3$-$s_4$ case and refer to appendix \ref{s1s1} and \cite{Taronna:2011kt} for more details.

\subsection{Generic dimension}\label{currexchd}
In this secttion we consider the computation of the flat space exchange amplitude for generic external spins and in arbitrary dimensions.

As in the previous Appendices we introduce the following building blocks for the most general cubic coupling:
\begin{equation}
f_{s_1,s_2,s_i}^{(k)}=Y_1^{s_1-k}Y_2^{s_2-k}Y_3^{s_3-k}\,G^k\,,
\end{equation}
where we use the same type of building blocks used in the 4d case:
\begin{subequations}
\begin{align}
Y_1&=\pl_{u_1}\cdot\pl_{x_2}\,,& Y_2&=-\pl_{u_2}\cdot\pl_{x_1}\,,& Y_{u}&=\tfrac12\pl_{u}\cdot\pl_{x_{12}}\,,\\
Y_3&=\pl_{u_3}\cdot\pl_{x_4}\,,& Y_4&=-\pl_{u_4}\cdot\pl_{x_3}\,,& Y_{v}&=\tfrac12\pl_{v}\cdot\pl_{x_{34}}\,,
\end{align}
\end{subequations}
together with
\begin{equation}
H_{ij}=\pl_{u_i}\cdot\pl_{u_j}\,,
\end{equation}
plus an additional building block $G$ which we define in two possible point splittings relevant to the exchange:
\begin{subequations}
\begin{align}
G_{u}&=Y_{1} H_{2u}+Y_2 H_{u1}+Y_{12} H_{12}\,,\\
G_{v}&=Y_{3} H_{4v}+Y_4 H_{v3}+Y_{34} H_{34}\,.
\end{align}
\end{subequations}
which is needed if we want to consider the most general cubic coupling in flat space. It is easy to show that the above structures define conserved currents with respect to the leg $i$.
The most general flat space structure then reads
\begin{equation}
\mathcal{V}_3=\sum_{s_1,s_2,s_3,k}g_{s_1,s_2,s_3}^{(k)}f_{s_1,s_2,s_3}^{(k)}\,.
\end{equation}

In order to compute the exchange we need to carry similar manipulations to those carried out in 4d. The propagator can be expressed in terms of a Gegenbauer polynomial and reads as
\begin{equation}
-\frac{i}{p^2}\,\mathcal{P}_r(u,v)\,,
\end{equation}
with the propagator numerator given by
\begin{equation}
\mathcal{P}_r(u,v)=\frac1{(r!)^2}\sum_{k=0}^{[r/2]}\alpha_k^{(r)}(u\cdot  v)^{s-2k}(u^2v^2)^k\,,
\end{equation}
and
\begin{equation}
\alpha_k^{(r)}=\frac{(-1)^k}{2^r}\frac{(\tfrac{d}{2}-2)_{r-k}}{(\tfrac{d}{2}-2)_{r}}\frac{r!}{k!(r-2k)!}\,,
\end{equation}
It is then clear that in order to evaluate the current exchange in flat space it is needed to compute the following contractions:
\begin{equation}
g_{s_1,s_2,s_i}^{(q)}g_{s_i,s_3,s_4}^{(q)}\underbrace{f_{s_1,s_2,s_i}^{(k)}f_{s_i,s_3,s_4}^{(q)}}_{\equiv F_{s_1,s_2|s_i|s_3,s_4}^{(k,q)}}\frac1{(r!)^2}\,(u\cdot  v)^{s-2k}(u^2v^2)^k\,.
\end{equation}
The above contractions can be rewritten in terms of differential operators acting on the cubic couplings and defined as:
\begin{subequations}
\begin{align}
\mathcal{O}_{u^2}&=-\left[Y_1 Y_2\pl_{Y_u}\pl_{G_u}+\frac{s}{2}\left(H_{12}^2\pl_{G_u}^2+2H_{12}\pl_{Y_u}\pl_{G_u}+\pl_{Y_u}^2\right)\right]\,,\\
\mathcal{O}_{v^2}&=-\left[Y_3 Y_4\pl_{Y_v}\pl_{G_v}+\frac{s}{2}\left(H_{34}^2\pl_{G_v}^2+2H_{34}\pl_{Y_v}\pl_{G_v}+\pl_{Y_v}^2\right)\right]\,,\\
\mathcal{O}_{u\cdot v}&=\Big[G_{12r}\cdot G_{r34}\pl_{G_u}\pl_{G_v}+\frac12 G_{12r}\cdot\pl_{x_{34}}\pl_{G_u}\pl_{Y_v}\nonumber\\&\hspace{100pt}+\frac12 G_{34r}\cdot\pl_{x_{12}}\pl_{G_v}\pl_{Y_u}+\left(\frac{t-u}{4}\right)\pl_{Y_u}\pl_{Y_v}\Big]\,,
\end{align}
\end{subequations}
with:
\begin{subequations}
\begin{align}
G_{12r}^\mu&\equiv \pl_{\pl_{u_r}}^\mu G_{12r}=\pl_{u_1}\cdot \pl_{x_2}\, \pl_{u_2}^\mu-\pl_{u_2}\cdot \pl_{x_1}\, \pl_{u_1}^\mu+\frac12\pl_{x_{12}}^\mu\,\pl_{u_1}\cdot\pl_{u_2}\,,\\
G_{r34}^\mu&\equiv \pl_{\pl_{u_r}}^\mu G_{r34}=\pl_{u_3}\cdot \pl_{x_4}\, \pl_{u_3}^\mu-\pl_{u_4}\cdot \pl_{x_3}\, \pl_{u_4}^\mu+\frac12\pl_{x_{34}}^\mu\,\pl_{u_3}\cdot\pl_{u_4}\,,
\end{align}
\end{subequations}
which enter the YM exchange.
We then get the following useful relations
\begin{subequations}
\begin{align}
F u^2\Big|_{u=0}&=\mathcal{O}_{u^2}\,F\,,\\
F v^2\Big|_{v=0}&=\mathcal{O}_{v^2}\,F\,,\\
F u\cdot v\Big|_{u,v=0}&=\mathcal{O}_{u\cdot v}F\,,
\end{align}
\end{subequations}
which allow to compute the exchange amplitude for arbitrary external excitations and couplings.

The explicit form of the above differential operators on a generic coupling can be computed easily using the multinomial expansion. In the following we focus for illustrative purpose on the $s$-$0$-$0$-$0$ case, as the general case can be derived with similar manipulations and is just more involved.

In this case none of the couplings depends on the $G$-structure and the trace and contraction operation simplify as:
\begin{align}
g_{s_1,0,r}^{(0)}g_{r,0,0}^{(0)}&Y_1^{s_1}Y_u^r Y_v^r\frac1{(r!)^2}\,(u\cdot  v)^{s-2k}(u^2v^2)^k\nonumber\\
&=\frac1{(r!)^2}g_{s_1,0,r}^{(0)}g_{r,0,0}^{(0)}\left[\left(\frac{t-u}{4}\right)\pl_{Y_u}\pl_{Y_v}\right]^{r-2k}\left[\frac{s}{2}\pl_{Y_u}^2\right]^k\left[\frac{s}{2}\pl_{Y_v}^2\right]^kY_1^{s_1}Y_u^r Y_v^r\nonumber\\
&=g_{s_1,0,r}^{(0)}g_{r,0,0}^{(0)}\left(\frac{t-u}{4}\right)^{r-2k}\left(\frac{s}{2}\right)^{2k}Y_{1}^{s_1}\,.
\end{align}
One can then easily carry on the resummation over spins in case the coupling converge factorially as $r$ goes to infinity. In 4d the above computation would have directly given the $k=0$ term plus local contributions, as the $s00$ current can be improved to a traceless one.

\section{More than one external spin}\label{s1s1}

As discussed in this note the problem to extract a local quartic coupling can be analised by studying Poncar\'e invariant amplitudes and imposing the cancellation of the $1/\Box$ non-localities when combining the homogeneous and particular solution to the Noether equations. Restricting for simplicity to 4d a cubic structure $f_{s_1s_1r}^{(s_1)}$ produces a numerator in the exchange amplitude given by $(G_{12r}\cdot p_3)^{s_1}$. A color ordered generating function for Poincar\'e invariant amplitudes with such numerators is given by:
\begin{equation}\label{genss00}
f_{ss00}=\sum_rg_{s_1s_100}^{[r]}\,\frac{({\sf t}/2)^{r-2s_1+1}}{{\sf su}}\,(\tfrac{{\sf su}}2\,G_{1100})^{s_1}\,,
\end{equation}
where we have defined
\begin{align}
G_{1100}&=\frac{1}{{\sf s}}\,G_{12r}\cdot p_3-\frac{1}{{\sf u}}\,Y_{14}\,Y_{23}\,,& [G_{1100},u_i\cdot p_i]&=0\,.
\end{align}
As before we can then proceed to extract the non-local part of \eqref{genss00} and impose that it is completely accounted by the current exchange \eqref{Exchs1s200}. A key difference with respect to the $s_1000$ case is that the corresponding pole in the ${\sf s}$-channel and in the ${\sf u}$-channel factorises on exchanges of different spin and hence require a fine-tuning for different coupling constants. Assuming that the exchanged spin is higher than the external spin: $r-s_1\geq0$ and focusing on the exchange accounting for the above numerator we have:
\begin{align}\label{ssnl2}
\mathcal{V}_{4\,\text{n.l.}}^{(ss00)}&=\frac{1}{{\sf s}}\,\left[(-1)^{s_1-1}g^{[r]}_{s_1s_100}+g^{(s_1)}_{s_1s_1r} \,g^{(0)}_{r00}\right]\,(G_{12r}\cdot p_3)^{s_1}\left(\tfrac{{\sf t}-{\sf u}}{4}\right)^{r-s_1}\,,\\
&+\frac1{{\sf u}}\,\left[(-1)^{s_1-1}g^{[r]}_{s_1s_100}+g^{(0)}_{s_1,r-s_1,0} \,g^{(0)}_{r-s_1,s_10}\right]\,(-\pl_{u_1}\cdot p_4)^{s_1}(\pl_{u_2}\cdot p_3)^{s_1}\left(\tfrac{{\sf t}-{\sf s}}{4}\right)^{r-s_1}\,.\nonumber
\end{align}
The above pole part should cancel to recover a local quartic coupling but this is equivalent to a non-trivial condition on the cubic coupling coefficients:
\begin{equation}
g^{(s_1)}_{s_1s_1r} \,g^{(0)}_{r00}=g^{(0)}_{s_1,r-s_1,0} \,g^{(0)}_{r-s_1,s_10}\,,
\end{equation}
which allows to solve for $g_{s_1s_100}^{[r]}=(-1)^{s_1}g^{(s_1)}_{s_1s_1r} \,g^{(0)}_{r00}$ achieving a maximal cancellation of the non-localities in eq.~\eqref{ssnl2}.
Considering the solution found by Metsaev:
\begin{equation}
g_{s_1s_2s_3}^{[k]}=\frac{(il)^{s_1+s_2+s_3-2k}}{\Gamma(s_1+s_2+s_3-2k)}\,,
\end{equation}
it is easy to see that the above condition is satisfied which implies that the above quartic couplings are local at least for exchanged spins big enough. In general the most general coupling constant satisfying the above condition is given by a general function $g_{s_1s_2s_3}^{[k]}=f(s_1+s_2+s_3-2k)$. That the function should be fixed to be the $\Gamma$-function is a consequence of locality for other choices of the external legs as discussed in section \ref{lightcone}.

Again, however, the above Poincar\'e invariant solution can correctly account for exchanges in the ${\sf s}$-channel such that $r>2s_1-1$. Therefore, the appearence of the non-local obstruction is a generic feature.

Similar result apply also to the following generating function of $s_1s_200$ amplitudes:
\begin{equation}\label{gens1s200}
f_{s_1s_200}=\sum_rg_{s_1s_200}^{[r]}\,\frac{({\sf t}/2)^{r-s_1-s_2+1}}{{\sf su}}\,(\tfrac{{\sf su}}2\,G_{1100})^{s_2}(\tfrac{{\sf su}}2\,G_{1000})^{s_1-s_2}\,,
\end{equation}
with
\begin{equation}
{\sf su}\,G_{1000}\equiv {\sf u} Y_{12}-{\sf s} Y_{14}\,.
\end{equation}
One can then convince himself that the above requirements are equivalent to those analysed in section \ref{lightcone}. It would be interesting to study the above maximal cancellation in higher-dimensions and investigate how the corresponding solution compares with \cite{Sleight:2016dba}.

\subsection{Trivially gauge invariant exchanges}\label{singleC}

It is useful to list, following \cite{Taronna:2011kt}, also the exchanges which do not give rise to non-local obstructions of the type discussed in this note.
As discussed in the previous sections there are indeed homogeneous solutions which are gauge invariant becuse of a cancellation between  two channels (referred to as open-string like amplitudes in \cite{Taronna:2011kt}), homogeneous solutions which are gauge invariant in combinations of 3 channels (referred to as closed-string like amplitudes in \cite{Taronna:2011kt}) and one should not forget the possibility of completing a \emph{single} exchange to a full gauge invariant amplitude with a proper choice of contact terms. As we will see almost all exchanges, but finitely many at fixed external spins, are of this type and for these reason they do not induce a priory such non-local obstructions. Non-local $1/\Box$ obstructions arise only when the propagating excitation has spin which is lower than some upper bound depending on the spin of the external polarisations.

This observation also motivates our convergence assumptions in the moment the tail of the sum over spin up to infinity can be shown to be free of any locality obstruction so that a finite number of obstruction arises for any given external spins and couplings.

Such exchanges are associated to the cubic couplings referred in \cite{Joung:2013nma} as \emph{class} I and \emph{class} II, which up to local field redefinitions are gauge invariant off-shell for at least two of the fields:
\begin{align}\label{offgauge}
    \delta_{\epsilon_i}^{(0)}(f_{s_1,s_2,s_3}^{[k]}+\Delta f_{s_1,s_2,s_3}^{[k]})&\equiv 0\,,& i&=1,2\,.
\end{align}
Above $\Delta f_{s_1,s_2,s_3}^{[k]}$ represents the appropriate improvement term which allows off-shell gauge invariance and whose existence has been studied\footnote{A convenient generating function form was given for all of them as:
\begin{align}
&\mathcal{K}^{(3)}(Y_i,H_{12},H_{23},H_{31})\,,& i&=1,2,3\,,
\end{align}
with
\begin{equation}
H_{ij}=\pl_{u_i}\cdot\pl_{u_j}\,\pl_{x_i}\cdot\pl_{x_j}-\pl_{u_i}\cdot\pl_{x_j}\pl_{u_i}\cdot\pl_{x_j}\,.
\end{equation}} in \cite{Joung:2013nma}
The simplest couplings of this type are $s$-$s$-$0$ couplings which satisfy the above condition when written in the curvature square form $R^{\mu(s)\nu(s)}R_{\mu(s)\nu(s)}\phi$. 

One can then compute the current exchange for such off-shell gauge invariant couplings keeping the off-shell gauge invariant legs as external legs:
\begin{equation}\label{homogeneral}
\mathcal{E}_{\sf s}=\left[f_{s_1,s_2,s_i}^{[k]}(\pl_{u_i})+\Delta f_{s_1,s_2,s_i}^{[k]}(\pl_{u_i})\right]\left[f_{s_3,s_4,s_i}^{[q]}(\pl_{w_i})+\Delta g_{s_3,s_4,s_i}^{[k]}(\pl_{w_i})\right]\mathcal{P}_{s_i}(u_i,w_i)\,.
\end{equation}
The corresponding exchange amplitude is then directly a solution to the homogeneous equation as a consequence of the off-shell gauge invariance of the cubic couplings external legs \eqref{offgauge}:
\begin{align}
\delta_{\epsilon_i}^{(0)} \mathcal{E}_{\sf s}&\equiv 0\,,& \forall&\quad i\,.
\end{align}
Subtracting to such homogeneous solutions the $1/\Box$ singularity which is proportional to the on-shell cubic couplings:
\begin{equation}
f_{s_1,s_2,s_3}^{[k]}(\pl_{u_i})\,f_{s_1,s_2,s_3}^{[q]}(\pl_{w_i})\,\mathcal{P}_{s_i}(u_i,w_i)\,.
\end{equation}
one is then left with the required local contact term which, as anticipated, exists in this cases at the level of the single exchange:
\begin{multline}
S^{(4)}=\left[f_{s_1,s_2,s_3}^{[k]}(\pl_{u_i})+\Delta f_{s_1,s_2,s_i}^{[k]}(\pl_{u_i})\right]\left[f_{s_1,s_2,s_3}^{[k]}(\pl_{w_i})+\Delta f_{s_1,s_2,s_i}^{[k]}(\pl_{w_i})\right]\mathcal{P}_{s_i}(u_i,w_i)\\-f_{s_1,s_2,s_3}^{[k]}(\pl_{u_i})\,f_{s_1,s_2,s_3}^{[k]}(\pl_{w_i})\,\mathcal{P}_{s_i}(u_i,w_i)\,.
\end{multline}
The latter is local by construction since it is just defined in this case as the contact part of the exchange amplitude.

In this way one generates an infinite number of gauge invariant homogeneous solutions \eqref{homogeneral} which being gauge invariant by themselves can be added to compensate as many non-localities as possible in the full contact solution $S^{(4)}$ and whose consistency can only be studied at quintic order. The general condition for \eqref{offgauge} to be satisfied is that the triangular inequality is violated which is possible if the exchanged spin is bounded from below.

This observation has an interesting corollary. Namely that any pole of the type:
\begin{equation}
\frac{({\sf t}-{\sf u})^{r}}{{\sf s}}\,(\pl_{u_1}\cdot p_2)^{s_1}(-\pl_{u_2}\cdot p_1)^{s_2}(\pl_{u_3}\cdot p_4)^{s_3}(-\pl_{u_4}\cdot p_3)^{s_4}\,,
\end{equation}
together with all analogous $1/\Box$ non-localities coming from the exchanges computed in appendix \ref{currexchd} can be formally compensated by an homogeneous solution if $r>\text{max}(s_1+s_2,s_3+s_4)$.

The latter bound can be improved for exchanges involving cubic couplings $f_{s_1,s_2,s_3}^{[k]}$ in \eqref{cub} with low values of $k$. 
The simplest example can be given for $k=0$:
\begin{equation}
f_{s_1,s_2,s_3}^{(0)}=Y_1^{s_1}Y_2^{s_2}Y_3^{s_3}\,.
\end{equation}
In this case \eqref{offgauge} can be obtained from the original coupling with a simple trick. Assuming $s_3$ to be the exchanged spin and $s_1\geq s_2$, one replaces the original cubic coupling as:
\begin{multline}
(\pl_{u_1}\cdot p_2)^{s_1}(-\pl_{u_2}\cdot p_1)^{s_2}(\pl_{u_3}\cdot p_1)^{s_1-s_2}\,Y_3^{s_3-(s_1-s_2)}\\\longrightarrow ([\pl_{u_1}p_1]^{s_1}\cdot[\pl_{u_2}p_2]^{s_2})^{\mu(s_1-s_2)}(\pl_{u_3}^{s_1-s_2})_{\mu(s_1-s_2)}\,Y_3^{s_3-(s_1-s_2)}\,,
\end{multline}
with
\begin{equation}
[\pl_{u_i} p_i]=(\pl_{u_i})_{\mu} (p_i)_{\nu}-(\pl_{u_i})_{\nu} (p_i)_{\mu}\,,
\end{equation}
and $\cdot$ simply meaning contraction of all naked indices. The above replacement is possible and can be shown to produce an on-shell equivalent coupling iff $s_3>s_1-s_2$. Such coupling is now gauge invariant off-shell with respect to the legs with spin $s_1$ and $s_2$ and the corresponding exchange amplitudes will directly give rise to the sought homogeneous solution for $r>\text{max}(|s_1-s_2|,|s_3-s_4|)$.

The above is just the simplest example of this class of cubic couplings studied in \cite{Joung:2013nma} and is the cubic analogue of the curl type couplings mentioned in section \ref{Homosol}. Let us stress that the above antisymmetrisation process only adds terms proportional to $p_1\cdot p_2$ which, when computing the exchange, will generate the sought local contact terms as $p_1\cdot p_2$ precisely cancels the propagator pole factor.

Let us conclude this appendix stressing the key limitation of the above type of homogeneous solution. While the reminder of the sum over spins in a given exchange amplitude is of this type the above homogeneous solution cannot compensate all $1/\Box$ contributions in the actual HS exchange. This is to be expected as otherwise there would be no non-trivial constraint on the cubic coupling coefficient. A theory with only type $I$ and $II$ couplings mentioned in this appendix would otherwise be consistent with Noether procedure but would not include non-abelian (gravitational) interactions or any non-abelian deformation of the linear gauge transformations. These indeed require gauge invariance of the homogeneous solution up to compensations among more than one single channel.

\end{appendix}

\bibliography{refs}
\bibliographystyle{JHEP}

\end{document}